\documentclass[%
 aip,
 amsmath,amssymb,
 reprint,%
 nofootinbib,
,floatfix]{revtex4-1}

\usepackage{graphicx}
\usepackage{dcolumn}
\usepackage{bm}

\usepackage{bbm}
\usepackage{braket}
\usepackage{stackengine} 
\stackMath 

\usepackage{mleftright} 
\usepackage{siunitx} 
\DeclareSIUnit{\cps}{cps} 
\usepackage[breaklinks]{hyperref}
\usepackage[all]{hypcap}    

\usepackage{xcolor}
\hypersetup{
    colorlinks = true,
    citecolor={blue},
    linkcolor={red},
    allbordercolors = {white},
    urlcolor={blue}
}
\usepackage{mathtools} 
\usepackage{IEEEtrantools} 
\usepackage{cleveref}
\usepackage{pm-isomath}

\let\diff\relax  
\usepackage{diffcoeff}
\diffdef{}{
long-var-wrap = dv,
op-symbol = \mathrm{d},
op-order-sep = 0 mu,
denom-term-sep = 3 mu,
}
\diffdef{p}{
long-var-wrap = dv,
denom-term-sep = 3 mu,
}

\newcommand{\IEEElabel}[1]{\addtocounter{equation}{-1}\refstepcounter{equation}\label{#1}}

\newcommand{\vs}[1]{\bm{#1}} 
\newcommand{\ms}[1]{\bm{#1}} 

\newcommand{\vo}[1]{\hat{\bm{#1}}}

\newcommand*{\dagg}{\ensuremath{^{\dagger}}}

\newcommand*{\bnum}[1]{\boldsymbol{#1}}

\newcommand{\idM}{\ensuremath{\openone}}
\newcommand{\idO}{\ensuremath{\hat \openone}}
\newcommand{\nullM}{\boldsymbol{0}} 

\newcommand*{\pii}{\ensuremath{\mathrm{\ISOpi}}} 
\newcommand*{\ii}{\ensuremath{\mathrm{i}}} 

\newcommand*\eh[1]{\mathop{}\!\mathrm{e}^{#1}}  

\newcommand*{\dd}[1][{}]{\dl3^{#1}}

\DeclareMathOperator{\diag}{diag}
\DeclareMathOperator{\tr}{tr}
\DeclareMathOperator{\sech}{sech} 

\let\Re\relax
\DeclareMathOperator{\Re}{Re} 

\let\Im\relax
\DeclareMathOperator{\Im}{Im} 

\newcommand*{\abs}[1]{\ensuremath{\mleft|{#1}\mright|}}
\newcommand*{\avs}[1]{\ensuremath{\abs{{#1}}^2}} 

\newcommand{\reals}{\mathbb{R}} 
\newcommand{\cplane}{\mathbb{C}} 

\newcommand{\aope}[1][{}]{\ensuremath{\hat a^{\vphantom{\dagger}{#1}}}}
\newcommand{\adag}[1][{}]{\ensuremath{\hat a^{\dagger {#1}}}}

\newcommand{\vac}{\ket{0}} 

\newcommand{\normOrd}[1]{\mathop{\vcentcolon}\nolimits\!#1\!\mathop{\vcentcolon}\nolimits}

\newcommand*{\tran}{^{\mkern-1.5mu\mathsf{T}}} 
\newcommand*{\trans}{^\mathsf{T}}              

\newcommand*{\mqty}[1]{\ensuremath{
\begin{pmatrix}
#1
\end{pmatrix}}}

\newcommand*{\amqty}[2]{\ensuremath{
\begin{pmatrix*}[#1]
#2
\end{pmatrix*}}}

\newcommand*{\vpdv}[3][]{\ensuremath{\bm{\partial}{#2}^{#1}_{#3}\,}}

\newcommand*{\eval}[1]{\ensuremath{\left. {#1} \right|}}

\newcommand*{\qq}[1]{\ensuremath{\quad \text{#1} \quad}}

\newcommand*{\vaope}[1][{}]{\ensuremath{\vs{\hat a}^{{#1}}}}
\newcommand*{\vadag}[1][{}]{\ensuremath{\vs{\hat a}^{\dagger {#1}}}}

\newcommand*{\ladvo}{\ensuremath{{\hat{\bm{\mathfrak a}}}}}
\newcommand*{\quadvo}{\ensuremath{{\hat{\bm{\mathfrak q}}}}}

\newcommand*{\refcite}[1]{Ref.~\onlinecite{#1}}

\makeatletter
\def\@email#1#2{%
 \endgroup
 \patchcmd{\titleblock@produce}
  {\frontmatter@RRAPformat}
  {\frontmatter@RRAPformat{\produce@RRAP{*#1\href{mailto:#2}{#2}}}\frontmatter@RRAPformat}
  {}{}
}%
\makeatother
\begin{document}


\title[Simulating the Photon Statistics of Multimode Gaussian States]{Simulating the Photon Statistics of Multimode Gaussian States by Automatic Differentiation of Generating Functions}


\author{Erik Fitzke}%
\affiliation{Institute for Applied Physics, Technische Universität Darmstadt,\\Schlossgartenstra\ss e 7,  64289 Darmstadt, Germany
}

\author{Florian Niederschuh}%
\affiliation{Institute for Applied Physics, Technische Universität Darmstadt,\\Schlossgartenstra\ss e 7,  64289 Darmstadt, Germany
}

\author{Thomas Walther}
\email{thomas.walther@physik.tu-darmstadt.de}
\affiliation{Institute for Applied Physics, Technische Universität Darmstadt,\\Schlossgartenstra\ss e 7,  64289 Darmstadt, Germany
}

\date{\today}

\begin{abstract}
Advances in photonics require photon-number resolved simulations of quantum optical experiments with Gaussian states. 
We demonstrate a simple and versatile method to simulate the photon statistics of general multimode Gaussian states. The derived generating functions enable simulations of the photon number distribution, cumulative probabilities, moments, and factorial moments of the photon statistics of Gaussian states as well as of multimode photon-added and photon-subtracted Gaussian states. Numerical results are obtained by automatic differentiation of these generating functions by employing the software framework PyTorch. Our approach is particularly well suited for practical simulations of the photon statistics of quantum optical experiments in realistic scenarios with low photon numbers, in which various sources of imperfections have to be taken into account. As an example, we calculate the detection probabilities for a recent multipartite time-bin coding quantum key distribution setup and compare them with the corresponding experimental values.
\end{abstract}


\maketitle

\section{Introduction}
\label{sec:introduction}
Recent progress in the generation, manipulation, and detection of photonic quantum states has led to new applications in the field of photonic quantum information processing and sensing.
An example is photonic quantum computing, which can be realized by using only single-photon sources, beam splitters, phase shifters, and photon detectors~\cite{Knill_2001}. Other applications such as quantum key distribution~(QKD) or quantum imaging have gained considerable attention over the last years and become commercially relevant, requiring the development of sophisticated optical setups working at the single-photon level~\cite{Xu_2020,Moreau_2019}.
Photon-number resolved~(PNR) detection of quantum states opens new pathways to experiments and applications requiring the simulation of such experiments. 

The strength of nonlinear optical interactions between light and matter typically decreases rapidly with the order of the nonlinear effect. Therefore, many common photonic quantum states are described by Hamiltonians that are at most quadratic in the creation and annihilation operators. States with such Hamiltonians are called Gaussian states~(GSs) and include e.g.\ vacuum, coherent states, squeezed states, thermal states, or states generated by spontaneous parametric down-conversion~(SPDC). Transformations by optical setups described by such Hamiltonians, introduced e.g.\ by phase shifters or beam splitters, map GSs to other GSs. Hence, the capability to simulate the photon statistics of GSs is of great practical relevance. The covariance formalism describes GSs by a covariance matrix and a displacement vector and allows the modeling of many common effects on GSs in experiments, such as losses, phase shifts, and interference at beam splitters, by relatively simple matrix transformations of the covariance matrix and displacement vector~\cite{Wang_2008_Quantum_information,Weedbrook_2012,Olivares_2012,Adesso_2014}. Therefore, it lends itself to the implementation of simulations of quantum optical experiments. Importantly, the covariance formalism, described briefly in \cref{sec:Covariance_Formalism}, carries the full information about the photon statistics of the state. 

While non-PNR detectors such as single-photon avalanche photodiodes are common, a variety of detector types capable of PNR detection have been developed as well~\cite{Divochiy_2008}. Transition edge sensors combine high detection efficiencies and PNR capabilities, but can only be operated at comparatively low count rates~\cite{Schmidt_2018}. Superconducting nanowire single-photon detectors (SNSPDs) can be operated at higher count rates and allow to realize PNR detection by evaluating the electronic output pulse shape depending on the number of incident photons~\cite{Cahall_2017} or by subdividing the light-sensitive area into multiple pixels~\cite{Tao_2019,Steinhauer_2021,He_2022,Cheng_2022}. SNSPD detectors achieving photon-number resolution with eight pixels are commercially available~\cite{id281_brochure}. Not only SNSPDs but also single-photon avalanche detectors have been multiplexed, in time~\cite{Achilles_2004,Kruse_2017} or space~\cite{Heilmann_2016}, to realize PNR detectors. In such multiplexing setups, multiple photons can hit the same sub-detector so that the photons are not resolved. But, with an increasing number of detectors, the counting statistics converge to the actual PND~\cite{Heilmann_2016,Sperling_2012_True_photocounting}.

The task to find the detection probability $p(\vs n) = p(n_1,\dots, n_S)$ for $n_1$~photons in mode~1, $n_2$~photons in mode~2 etc.\ of a GS with $S$~modes, i.e.\ its photon number distribution~(PND), is known as the Gaussian boson sampling~(GBS) problem~\cite{Hamilton_2017}. Experimentally, GBS has been realized e.g.\ using networks of beam splitters on photonic chips~\cite{Tillmann2013,Broome_2013}.
Large-scale GBS experiments have been realized for example in the context of quantum computing and to pursue the demonstration of the computational advantage of quantum computers over classical computers~\cite{Brod_2019, Zhong_2020_Quantum, Zhong_2021_Phase}. \refcite{Bromley_2020} discusses possible applications for GBS-based quantum computing. The computational complexity of simulating GBS has been investigated as a benchmark for optical quantum computing comparing it to GBS simulations on classical computers~\cite{Aaronson_2011, Lund_2014, Rahimi_2015, Hamilton_2017, Quesada_2018, Kruse_2019, Quesada_2022_Quadratic, Bjoerklund_2019, Bulmer_2022}.
The fact that GBS is investigated in the context of computational quantum supremacy indicates that calculating the solution to GBS problems can require substantial computational resources.
The PND can be calculated by evaluating expressions involving matrix functions called the Hafnian and loop Hafnian for PNR detection as well as of functions called the Torontonian and loop Torontonian for non\nobreakdash-PNR detection~\cite{Hamilton_2017,Quesada_2018,Bjoerklund_2019,
Quesada_2019_FranckCondon,Bulmer_2022_Threshold}. The operation count for the evaluation of these functions scales exponentially with the number of detected photons and optimization of the algorithms is an active field of research~\cite{Bulmer_2022}.

The previously mentioned methods for GBS computations have been designed for calculating PNDs with many photons efficiently, but do not offer much flexibility. Typically, quantum optical setups not designed for the particular task of GBS quantum computation are operated at low photon numbers. Examples are applications in quantum ghost imaging and QKD, where coincidences involving only a relatively small number of detectors are relevant. If such a setup is to be simulated, minimizing the required computational resources is not necessarily the main concern. However, often a simulation method is required that is flexible enough to include imperfections inevitably present in real setups. In this paper, we present such a simulation method. The purpose of our method is not to compete with specialized GBS algorithms in terms of performance, but rather to provide a flexible and simple way to compute the photon statistics for imperfect setups. 

Two relevant effects in experimental setups are the simultaneous detection of multiple modes by the same detector and noise in the detection process. These effects require the convolution of probability distributions which can conveniently be expressed by the multiplication of the corresponding probability-generating functions.
Therefore, we derive generating functions of the photon statistics in terms of the covariance matrix and displacement vector in \cref{sec:detection_of_Gaussian_states}. 
Our generating-function approach is an alternative to the established expressions for the PND involving Hafnian\nobreakdash-type functions. This different perspective on the GBS problem allows for deriving several related expressions for the generating functions of cumulative probabilities, raw or central moments, and rising or falling factorial moments of the detection statistics in \cref{ssec:Generating_function_from_Covariance}, for which so far no systematic method existed. Furthermore, our method allows us to calculate the same quantities for certain non-Gaussian states called photon-added and photon-subtracted GSs as shown in \cref{ssec:nonGaussian_states} considerably extending the range of possible applications.

Generating functions need to be differentiated repeatedly to retrieve probabilities or moments. We show that these derivatives can be evaluated numerically by automatic differentiation~(AD) without much effort. An advantage of AD is that it provides accurate numerical results while hiding the whole complexity of the calculation from the user. We discuss the usage of AD for our application in \cref{ssec:AD} and provide an implementation example in \refcite{Fitzke2022_Pytorch_Technical_Report}.

Our simulation method consists of two steps, connected by the generating function. First, the quantum state and the optical setup are modeled in the covariance formalism. Second, the photon statistics are obtained by AD of the generating function expressed in terms of the covariance matrix and displacement vector. Both steps can be easily implemented with widely available software, making the method very practical.
However, it is expected that optimized algorithms for the evaluation of Hafnian-type functions will outperform general-purpose AD algorithms for the particular task of calculating the PND. To show that our method can nevertheless be used to simulate various aspects of Gaussian photon statistics efficiently for small photon numbers, we apply it to multiple examples with common GSs in \cref{sec:Application_to_common_Gaussian_states}.

As a more complex application, we present in \cref{sec:Application_QKD_System} a simulation of an entanglement-based QKD system we presented recently in~\refcite{Fitzke_2022_Scalable}. It demonstrates the strengths of our method to readily incorporate relevant imperfections such as noise, detection efficiencies, and simultaneous detection of multiple modes. Although the experimental setup is complex, the simulation is straightforward because it can be broken down into elementary operations described by basic Gaussian state transformations. The simulation results are in very good agreement with the experimental values. Furthermore, we show that the contribution of multi-photon pair emission to the quantum bit error rate of the QKD system can be estimated by applying Bayes' theorem to PNR simulation results. 

\section{Generating functions for the detection statistics of Gaussian states}
\label{sec:detection_of_Gaussian_states}

Our method to simulate the photon statistics of GSs uses generating functions for the photon statistics in terms of the covariance matrix~$\ms \Gamma$ and the displacement vector~$\vs d$. They connect the covariance formalism to the photon statistics and are therefore essential for our simulation method.
In \cref{ssec:Generating_functions} we briefly summarize the relevant properties of generating functions for probability distributions. In the following \cref{ssec:Generating_operators}, we briefly recap the formulation of the PND in terms of a generating operator and extend it to generating operators for the moments and factorial moments. 

\subsection{Generating operators for photon detection probabilities and moments} 
\label{ssec:Generating_operators}

Assuming a photon detection process where photons in a single mode are detected independently of each other with an efficiency~$\eta$, the probability to detect $n$~photons from a quantum state~$\rho$ is given by~\cite{Kelley_1964, Lax_1973, Mandel_Wolf_1995, Walls2008}
\begin{equation}\label{eq:quantum_mandel_formula}
p(n) = \biggl\langle\normOrd{\frac{(\eta \hat N )^n}{n!}\eh{-\eta \hat N}}\biggr\rangle\,.
\end{equation}
Here, normal order is indicated by~$\normOrd{\:\:}$ and the photon number operator is $\hat N = \adag \aope$. By inserting \cref{eq:quantum_mandel_formula} into the defining equation of a probability generating function~$h(y)$ (cf.~\cref{eq:PGF}), the PND can be obtained from the expectation value of a generating operator $\hat h(y)$~\cite{Kelley_1964, Lax_1973, Mandel_Wolf_1995, Walls2008}. This method has recently been used by Nauth to express the PND
of biphoton states in terms of the covariance matrix~\cite{Nauth_2022}. We extend this procedure to obtain similar generating operators for the moment generating function~$M(\mu,y)$ and rising factorial moment generating function~$R(y)$ as described in \cref{ssec:Generating_functions}:
\begin{IEEEeqnarray}{rCl}
\hat h(y) &=& \normOrd{\eh{\eta(y-1)\hat N}} \IEEElabel{eq:SM_prob_gen_operator}\,,\\
\hat M(\mu, y) &=& \eh{-y\mu} \normOrd{\exp\bigl(\eta (\eh y -1)\hat N\bigl)\,}\,,
 \IEEEeqnarraynumspace\IEEElabel{eq:SM_mom_gen_operator}\\
\hat R(y) &=& \frac{1}{1-y} \normOrd{\eh{\eta y \hat N /(1-y)}}\,.\IEEEeqnarraynumspace\IEEElabel{eq:SM_fall_fac_gen_operator}
\end{IEEEeqnarray}

These operators can be modified to take into account noise in the detection process. For that, according to the multiplication rule for generating functions (cf.\ \cref{ssec:Generating_functions}), the generating operator is simply multiplied by the generating function of the noise process. As an example, we consider noise with Poissonian statistics $p\ped{noise}(n) = \eh{-\nu}\nu^n/n!$ and noise parameter~$\nu$. The probability generating function of the noise is given by $h\ped{noise}(y) = \eh{\nu(y-1)}$ so that the generating operators from \crefrange{eq:SM_prob_gen_operator}{eq:SM_fall_fac_gen_operator} including noise read 
$\eh{\nu(y -1)} \hat h(y)$, $\exp\bigl(\nu(\eh y -1)\bigr)\hat M(\mu, y)$ and $\eh{\nu y/(1-y)} \hat R(y)$.

Similarly, noise with different statistics could be taken into account by multiplying the generating operators from \crefrange{eq:SM_prob_gen_operator}{eq:SM_fall_fac_gen_operator} with the generating function of the respective noise process. Another option is to include noise in the covariance formalism. Thermal noise can be modeled by coupling in a thermal state with a beam splitter or a matrix can be added to the covariance matrix to represent e.g.\ classical Gaussian noise or noise from amplification~\cite{Eisert_2007}.

\Crefrange{eq:SM_prob_gen_operator}{eq:SM_fall_fac_gen_operator} have in common that they involve special cases of the generating operator $\hat g(w(y)) = \normOrd{\exp(-w(y) \adag \aope)}$ for different functions~$w(y)$.

In \refcite{Adam_1995,Perina_1991} the operator~$\hat g$ has been extended to 
\begin{equation}\label{eq:general_generating_operator}
    \hat g(u, v, w) = \normOrd{\exp(u \aope +v \adag -w\adag \aope )}\,
\end{equation}
and has been related to the density matrix elements of~$\hat \rho$ by introducing $u$~and~$v$.

\subsection{Generating functions in terms of the covariance matrix and displacement vector}
\label{ssec:Generating_function_from_Covariance}

In experiments, often multiple modes, e.g.\ different polarization directions or frequency modes, enter the same detector. For example, calculating the simultaneous detection of multiple modes is required for modeling imperfect interference due to a mode mismatch with the model from \refcite{Takeoka_2015}, which we use in \cref{sec:Application_QKD_System}.
Furthermore, often joint detection probabilities between multiple detectors such as coincidence probabilities are of interest. Therefore, we generalize the calculation to multi-mode states. 

In \cref{app:Gen_fun_in_terms_of_covmat_and_disvec} we show that for multiple modes with vectors $\vs u$, $\vs v$, and $\vs w$, $G(\vs u,\vs v, \vs w) = \braket{\hat g(\vs u, \vs v, \vs w)}$ can be expressed in terms of the covariance matrix~$\ms \Gamma$ and displacement vector~$\vs d$ by 
\begin{IEEEeqnarray}{rCl}
G(\vs u, \vs v, \vs w) &=& \frac{\exp(-\vs z\trans \ms \Lambda^{-1} \ms W\vs z/2 + Z)}{\sqrt{\det \ms \Lambda}}\IEEElabel{eq:multi_photon_generating_function}
\end{IEEEeqnarray} with the diagonal matrix $\ms W =  \diag(\vs w) \oplus \diag(\vs w)$, $\ms \Lambda  = \idM + \ms W (\ms \Gamma -\idM)/ 2$, $Z = \sum_{s} u_s v_s w^{-1}_s$ and  $ \vs z = \vs d + \vs \zeta$ with $\vs \zeta$ as defined in \cref{eq:zeta_def}.
\Cref{eq:multi_photon_generating_function} is the generating function for the multivariate photon statistics from which the probabilities and moments are retrieved by repeated differentiation w.r.t.\ $D$~parameters $y_1\dots y_D$ for the detectors~$d=1 \dots D$, detecting $M_d$~modes, with additional Poissonian noise\footnote{Similarly, the generating function of noise processes with a different, non-Poissonian statistics could be taken into account by multiplying with the corresponding generating function.}~$\nu_d$. The total number of modes is $S = \sum_d M_d $ and the index~$s$ runs over all mode indices, i.e.\ enumerates~$m_d=1_d\dots M_d$ for all $D$~detectors in the order $1_1, 2_1, \dots, M_1, 1_2,\dots M_2,\dots \dots M_D$. Abbreviating $G(\vs w) = G(\bnum{0}, \bnum{0}, \vs w)$ and employing a multi-index notation\footnote{For tuples~$\vs x$ and~$\vs k$ we write $\vs x^{\vs k} = \prod_i x_i^{k_i}$, so that e.g.\ $\vs w^{\bnum{-1}} = \prod_s w_s^{-1}$. We also use 
$\vs n! = \prod_d n_d!$ and $\vpdv[\vs n]{}{\vs y} = \prod_d \big(\partial ^{n_d}/\partial y_d^{n_d}\big)$.
} we obtain:

\begin{widetext}
\begin{IEEEeqnarray}{lCl"s?L}
p(\vs n, \vs \nu, \vs \eta) = \braket{\hat \Pi_{\vs N= \vs n}} &=& \frac{1}{\vs n!}\vpdv[\vs n]{}{\vs y}\exp\Bigl(\sum\nolimits_d (y_d-1)\nu_d\Bigr) G(\vs w) \Big\rvert_{\abovebaseline[1pt]{\scriptstyle \vs y = \bnum 0}} & with & w_s = \eta_{m_d} (1-y_d)\,,\IEEEeqnarraynumspace\IEEElabel{eq:multi_photon_detection_probability}\\
p(\vs N \le \vs n, \vs \nu, \vs \eta) = \braket{\hat \Pi_{\vs N\le \vs n}} &=& \frac{1}{\vs n!}\vpdv[\vs n]{}{\vs y}(\bnum{ 1}-\vs y)^{\bnum{-1}}\exp\Bigl(\sum\nolimits_d (y_d-1)\nu_d\Bigr) G(\vs w)\Big\rvert_{\abovebaseline[1pt]{\scriptstyle \vs y = \bnum 0}} & with & w_s = \eta_{m_d} (1-y_d)\,, \IEEElabel{eq:multi_photon_cumulative_probability}\\
{\cal M}(\vs \mu, \vs k, \vs \nu, \vs \eta) = \braket{(\hat{\vs N}-\vs \mu)^{\vs k}} 
&=& \vpdv[\vs k]{}{\vs y}
\exp\Bigl(\sum\nolimits_d (\eh{y_d} -1)\nu_d - \mu_d y_d\Bigr)
G(\vs w)\Big\rvert_{\abovebaseline[1pt]{\scriptstyle \vs y = \bnum 0}} & with & w_s = \eta_{m_d} (1-\eh{y_d})\,,  \IEEElabel{eq:multi_photon_moments}\\
n_{(\vs k)}(\vs \nu, \vs \eta) = 
\Braket{\prod\nolimits_d\hat N_{d(k_d)}}
&=& \vpdv[\vs k]{}{\vs y} \exp\Bigl(\sum\nolimits_d y_d\nu_d\Bigr)G(\vs w) \Big\rvert_{\abovebaseline[1pt]{\scriptstyle \vs y = \bnum 0}} & with  &  w_s = -\eta_{m_d} y_d\,, \IEEElabel{eq:multi_photon_falling_factorial_moments}\\
n^{(\vs k)}(\vs \nu, \vs \eta) 
= \Braket{\prod\nolimits_d \hat N_d^{(k_d)}}
&=& \vpdv[\vs k]{}{\vs y} (\bnum{1}-\vs y)^{\bnum{-1}} \exp\biggl(\sum\nolimits_d \frac{\nu_d y_d}{1-y_d}\biggr) G(\vs w) \bigg\rvert_{\abovebaseline[1pt]{\scriptstyle \vs y = \bnum 0}} & with & w_s = \frac{\eta_{m_d}y_d}{y_d-1}\,. \IEEElabel{eq:multi_photon_rising_factorial_moments}
\end{IEEEeqnarray}
\end{widetext}

\Crefrange{eq:multi_photon_generating_function}{eq:multi_photon_rising_factorial_moments} connect the covariance formalism and the photon statistics by repeated derivatives and are therefore crucial for our simulation method.
Alternative expressions to calculate~$p(\vs n)$, restricted to the noiseless case, can be found in the literature and are discussed and compared to \cref{eq:multi_photon_detection_probability} in \cref{sec:discussion}. 
Our generating-function approach additionally yields \crefrange{eq:multi_photon_cumulative_probability}{eq:multi_photon_rising_factorial_moments} for the cumulative probabilities, raw moments (for~$\mu=0$), central moments (for~$\mu =\braket{\hat N}$), and rising and falling factorial moments directly incorporating the detection efficiency and noise. For general multi-mode GSs, to the best of our knowledge, expressions for these quantities have so far not been presented in the literature.

The cumulative probabilities are useful when ranges of photon numbers are to be analyzed jointly, especially when detection events in multiple detectors are considered. For example, the calculation of the coincidence probability to detect~$\le n_1$ photons in one detector and~$\le n_2$ in a second detector from the PND requires the evaluation of the probabilities for $(n_1+1)(n_2+1)$~combinations of photon numbers, but only one evaluation of \cref{eq:multi_photon_cumulative_probability}. Mean and variance are two of the most important moments of the PND and can be directly calculated from \cref{eq:multi_photon_moments}. The factorial moments are useful for the calculation of photon statistics of photon-added and photon-subtracted states (cf.~\cref{ssec:nonGaussian_states}). 

By using a formulation with detection operators~$\hat \Pi$, more complex detection events involving multiple detectors can be calculated. For example, the probability for the detection of $n_A$~photons in detector~$A$ and $n_B$~photons in detector~$B$ is given by $\braket{\hat \Pi_{N_A=n_B,N_B=n_B}} = \braket{\hat \Pi_{N_A=n_A}\hat \Pi_{N_B=n_B}}$ and the probability for~$n_1$ or~$n_2$ photons in the same detector with~$n_1 \neq n_2$ is given by $\braket{\hat \Pi_{N=n_1 \text{ or }N=n_2}} = \braket{\hat \Pi_{N=n_1}+\hat \Pi_{N=n_2}}$. 
Operators for complementary events can be defined as well, enabling the calculation of the probability to detect any photon number except for $n$~or for more than $n$~photons with $\hat \Pi_{N\neq n} = \idO -\hat \Pi_{N=n}$  and $\hat \Pi_{N>n} = \idO - \hat \Pi_{N\le n}$. The detection operators for different modes are then joined before the generating function is evaluated. This procedure was used for example in~\refcite{Takeoka_2015} to relate the detection probabilities for non\nobreakdash-PNR detectors to the detection of vacuum by considering operators~$\hat \Pi_{N>0} = \idO - \hat \Pi_{N=0}$. We extend this method to PNR detection. 
As an example, consider the probability to not detect  $n_A$~photons in detector~$A$ and to detect more than $n_B$~photons in detector~$B$:
\begin{IEEEeqnarray}{rCl}
\IEEEeqnarraymulticol{3}{l}{
\big\langle\hat \Pi_{N_A \neq n_A}\hat \Pi_{N_B> n_B}\big\rangle = \big\langle(\idO -\hat \Pi_{N_A = n_A})(\idO -\hat \Pi_{N_B\le n_B})\big\rangle}\nonumber\\
&=& 1 - \braket{\hat \Pi_{N_A = n_A}}-\braket{\hat \Pi_{N_B\le n_B}} + \braket{\hat \Pi_{N_A = n_A, N_B\le n_B}}\IEEEeqnarraynumspace
\end{IEEEeqnarray} 
The last term, for example, is given by
\begin{equation}
\diffp*[n_A,n_B]{\frac{\exp(-\vs d_{AB}\trans \ms \Lambda_{AB}^{-1} \ms W_{AB}\vs d_{AB}/2)}{n_A!n_B!(1-y_B)\sqrt{\det \ms \Lambda_{AB}}}}{y_A,y_B}\bigg \rvert_{\abovebaseline[1pt]{\substack{y_A= 0\\y_B = 0}}}\,,
\end{equation} where the modes for $A$~and~$B$ are kept in $\ms \Lambda$~and~$\vs d$.

Finally, we derive multivariate expressions for matrix elements of $\hat \rho$. In \cref{app:Matrix_elemnts_singlemode} we present a simple derivation of the matrix elements in terms of $g(u, v, w) = \braket{\hat g(u, v, w)}$. We generalize the single-mode expressions from \cref{eq:coherent_state_matrix_element,eq:closeby_matrix_elements} to the multimode case by using \cref{eq:multi_photon_generating_function}: 
\begin{IEEEeqnarray}{rCl}
\braket{\vs \alpha|\hat \rho|\vs \beta} & = & \eh{-(\avs{\vs \alpha}+\avs{\vs \beta})/2} G(\vs \alpha^*, \vs \beta, \bnum{1})\IEEElabel{eq:multi_coherent_matrix_element}\,,\\
\braket{\vs n|\hat \rho|\vs m} &=& \frac{(\bnum{-1})^{\vs l}}{\sqrt{ \vs n !\vs m !}}\vpdv[\vs l]{}{\vs w}\vpdv[\vs{\Delta_n}]{}{\vs u}\vpdv[\vs{\Delta_m}]{}{\vs v}  G(\vs u, \vs v, \vs w)\bigg \rvert_{\abovebaseline[1pt]{\substack{\vs u=\bnum 0\\\vs v=\bnum 0\\\vs w=\bnum 1}}}\,.\IEEEeqnarraynumspace\IEEElabel{eq:multi_closeby_matrix_elements}
\end{IEEEeqnarray}
In \cref{eq:multi_closeby_matrix_elements} we define $\vs l$ by $l_s = \min(n_s,m_s)$ as well as $\vs{\Delta n} = \vs n - \vs l$ and $\vs{\Delta m} = \vs m - \vs l$. 
An equation similar to \cref{eq:multi_closeby_matrix_elements} with $\vs l = \bnum 0$, $\vs{\Delta n} = \vs n$ and $\vs{\Delta m} = \vs m$ can be derived differently~\cite{Quesada_2019_Simulating} and can be related to an expression involving the loop Hafnian function. Compared to this expression, our \cref{eq:multi_closeby_matrix_elements} has the advantage that for each mode the number of derivatives to be evaluated is reduced from $n_s+m_s$ to $\max(n_s,m_s)$. When the derivatives are evaluated numerically, for example by automatic differentiation~(AD), our expression is especially advantageous for close-by values of $n_s$~and~$m_s$. 

\subsection{Photon statistics of non-Gaussian states derived from Gaussian states}
\label{ssec:nonGaussian_states}

The range of possible applications for our simulation method of the photon statistics can be extended by adapting it to apply to certain non-Gaussian states.

A special class of states derived from GSs~$\hat \rho$ are photon-added GSs~$\hat \rho_{+\vs k}$ and photon-subtracted GSs $\hat \rho_{-\vs k}$~\cite{Barnett_2018},
\begin{equation}
\hat \rho_{+\vs k} = \frac{\vadag[\vs k]\hat \rho \vaope[\vs k]}{m^{(\vs k)}} \qq{and}
\hat \rho_{-\vs k} = \frac{\vaope[\vs k]\hat \rho \vadag[\vs k]}{m_{(\vs k)}}\,,
\end{equation}
with $\vs k = k_1,k_2, \dots$~photons added or subtracted from the modes~$1,2,\dots$.
Here, the normally and anti-normally ordered moments~\cite{Adam_1995}
 $m_{(\vs k)}=
\tr(\vaope[\vs k]\hat \rho \vadag[\vs k]) = \braket{\vadag[\vs k]\vaope[\vs k]}$ and $m^{(\vs k)} =\tr(\vadag[\vs k]\hat \rho \vaope[\vs k]) = \braket{\vaope[\vs k] \vadag[\vs k]}$ ensure normalization. They can be obtained from the falling and rising factorial moments of the GS by setting $\eta = 1$~and~$\nu = 0$:
\begin{IEEEeqnarray}{rCcl}
m_{(\vs k)} &=& \sum_{\vs n\geq \vs k} &\braket{\vs n|\hat \rho|\vs n} \frac{\vs n!}{(\vs n-\vs k)!} = n_{(\vs k)}( \vs \nu =\bnum 0, \vs \eta = \bnum 1)\,. \IEEEeqnarraynumspace
\end{IEEEeqnarray}
Similarly, we directly obtain $m^{(\vs k)} = n^{(\vs k)}( \vs \nu =\bnum 0, \vs \eta = \bnum 1)$. Here, our \cref{eq:multi_photon_falling_factorial_moments,eq:multi_photon_rising_factorial_moments} for factorial moments are helpful because they allow to calculate $n_{(\vs k)}$~and~$n^{(\vs k)}$ directly. 

Photon addition and photon subtraction can have non-trivial effects on the PND of quantum states. An example is photon subtraction from a thermal state, which \emph{increases} the mean photon number of the state~\cite{Bogdanov_2017, Barnett_2018}. Photon-subtracted states can be approximately realized by inserting a beam splitter with high transmission into the beam and by conditioning the detection of the transmitted quantum state on the detection of a reflected photon~\cite{Bogdanov_2017, Barnett_2018, Zavatta_2008}. Photon addition has been realized by seeding an SPDC process with low gain so that the SPDC signal modes are emitted into the mode of the incident state. The generation of the photon-added state is then conditioned on the detection of an SPDC idler photon~\cite{Zavatta_2004, Parigi_2009}. 
Photon addition and subtraction can be used for example to enhance the signal-to-noise ratio in quantum ghost imaging~\cite{Liu_2021_Ghost}.

The matrix elements of photon-added and photon-subtracted GSs can be directly calculated by using $\aope[k] \ket l = \sqrt{l!/(l-k)!}\ket{l-k}$, given that~$l\geq k$, and $\adag[k] \ket l = \sqrt{(k+l)!/l}\ket{k+l}$:
\begin{IEEEeqnarray}{rCl}
 \braket{\vs  n|\hat \rho_{-\vs  k}|\vs l} &=& \frac{\braket{\vs n+\vs k|\hat \rho|\vs l+\vs k}}{ m_{(\vs k)}} \sqrt{\frac{(\vs l+\vs k)!(\vs n+\vs k)!}{\vs l!\vs n!}}\,,\IEEEeqnarraynumspace\\
    \braket{\vs n|\hat \rho_{+\vs k}|\vs l} &= &\frac{\braket{\vs n-\vs k|\hat \rho|\vs l-\vs k}}{ m^{(\vs k)}} \sqrt{\frac{\vs l!\vs n!}{(\vs l-\vs k)!(\vs n-\vs k)!}}\,.
\end{IEEEeqnarray}
The expressions for $\braket{\vs n+\vs k|\hat \rho|\vs l+\vs k}$ and $\braket{\vs n-\vs k|\hat \rho|\vs l-\vs k}$ can be calculated from the GS's matrix elements by using 
\cref{eq:multi_closeby_matrix_elements}.

To derive explicit expressions for the photon statistics, we first consider the single mode case of~$\hat \rho_{-k}$. The operator $ \adag[k] \hat g \aope[k]$ is in normal order and  we obtain from the optical equivalence theorem in~\cref{eq:optical_equivalence_theorem}
\begin{equation}
\tr(\hat \rho \adag[k] \hat g\aope[k]) = \int_\cplane P(\alpha)\abs{\alpha}^{2k}\eh{u \alpha + v\alpha^* - w\avs{\alpha}}\dd[2]\alpha\,.
\end{equation}
By comparison to the phase-space representation of~$\hat g$ from \cref{eq:phase_space_representation_of_g}, the expectation value of~$\hat g$ with respect to~$\hat \rho_{-\vs k}$ can be expressed in terms of the generating function of the underlying GS~$\hat \rho$: 
\begin{equation}\label{eq:photon_subtracted_generating_function}
\tr(\hat \rho_{-\vs k}\hat g(\vs u, \vs v, \vs w)) =\frac{(\bnum{-1})^{\vs k}}{m_{(\vs k)}}\vpdv[\vs k]{}{\vs w} G(\vs u, \vs v, \vs w)\,.
\end{equation}

For the photon-added GS, we consider for simplicity only~$G(\vs w)$, as $\vs u$~and~$\vs v$ are essentially only required for the calculation of the matrix elements discussed above.
In \cref{app:photon_added_derivation}, we derive the expression for the single-mode case which generalizes to
\begin{IEEEeqnarray}{rCl}
\tr(\hat \rho_{+\vs k}\hat g(\vs w))
&=& \frac{1}{m^{(\vs k)}} \vpdv[\vs k]{}{\vs r} \eval{G(\tilde{\vs w}) \prod_s \frac{1}{1-r_s(1-w_s)} }_{\abovebaseline[1pt]{\scriptstyle{\vs r=\bnum 0}}}\IEEElabel{eq:photon_added_generating_function}
\end{IEEEeqnarray} with~$\tilde w_s = 1 -[(1- w_s)^{-1} -r_s]^{-1}$.

This means that all the quantities of the photon statistics for which we have derived expressions above can also be calculated for multi-mode photon-added and photon-subtracted GSs. In this context, $G(\vs w) = \tr(\hat \rho\hat g(\vs w))$ in
\crefrange{eq:multi_photon_detection_probability}{eq:multi_photon_rising_factorial_moments} is replaced by the expressions for $\tr(\hat \rho_{-\vs k}\hat g(\vs w))$ and $\tr(\hat \rho_{+\vs k}\hat g(\vs w))$ 
from \cref{eq:photon_subtracted_generating_function,eq:photon_added_generating_function}.

To our knowledge, expressions for the PND of photon-added and photon-subtracted GSs have so far only been reported for single-mode states~\cite{Xu_2010,Wang_2012_Photon_subtracted,Wang_2012_Photon_number,Ghiu_2013,Ghiu_2014}. Our expressions enable the calculation of the photon statistics for the general multi-mode case and additionally enable the calculation of moments, factorial moments, and matrix elements. 
Moreover, the expressions are evaluated by repeated differentiation and can therefore be calculated with the same tools as the expressions for GSs.

Certain non-Gaussian states~$\hat \rho'$, such as NOON states and cat states, can be generated from larger GSs~$\hat \rho$ by PNR detection~\cite{Zhai_2013, Daiqin_2019,Quesada_2019_Simulating,Gagatsos_2019}. The effects of the optical setup on such states can be modeled using the covariance formalism as described in~\refcite{Daiqin_2019}.
Consider a GS~$\hat \rho$ with $(M'+M)$~modes, of which $M$~modes are detected with PNR detectors so that a state~$\hat \rho'$ with $M'$~modes remains. By conditioning the further analysis of~$\hat \rho'$ on the detected PND, the non-Gaussian states are produced probabilistically~\cite{Daiqin_2019}. Unitary transformations~$U'$ acting on~$\hat \rho'$ can simply be modeled by applying the unitary transformation~$U = U'\otimes \idO_{M\times M}$ to~$\hat \rho$~\cite{Daiqin_2019}.
For the simulation, the required PNR detection pattern on the $M$~modes is fixed and the photon detection pattern of interest is applied to the remaining $M'$~modes. Unitary transformations that are linear or quadratic in $\aope$~and~$\adag$ can be applied to~$\hat \rho$ in the covariance matrix formalism before the detection~\cite{Daiqin_2019}. 
By using this procedure from \refcite{Daiqin_2019}, our method can be directly applied to simulate states that can be obtained from GSs via PNR detection.

To summarize, we note that our method to simulate the photon statistics of GSs can also be applied to photon-added and photon-subtracted GSs as well as to non-Gaussian states derived from larger GSs by PNR detection. The class of states that can be simulated and the range of possible applications of the simulation method are thereby greatly extended. 

\subsection{Automatic differentiation for retrieval of probabilities and moments}
\label{ssec:AD}

Multiple options exist to evaluate the derivatives of the generating functions. One option used in \refcite{Thomas_2021} is to use finite-difference approximations, but this method can accumulate numerical inaccuracies. Another option to retrieve probabilities from the generating function is to approximate Cauchy's integral formula on a circle around the origin in the complex plane.
In \refcite{Abate_1992_Numerical} this method has been applied to probability generating functions and error bounds for the approximations have been given. It is extensively discussed in \refcite{Abate_1992_Fourier} and expressions for moments and multi-dimensional distributions are given in Refs.~\onlinecite{Choudhury_1994, Choudhury_1996_Computing, Choudhury_1996_Numerical}.

A versatile method to evaluate derivatives of functions numerically is automatic differentiation~(AD)~\cite{Griewank_2008, Naumann_2012}.
AD breaks down a function to be differentiated into elementary operations such as addition, multiplication, or~$\sin(x)$. The differentiation rules for these operations are then applied and combined according to the chain rule to the derivative of the overall function, which is finally evaluated. For example, instead of approximating the derivative of~$\sin(x)$ numerically, AD uses the fact that the derivative is given by~$\cos(x)$ and evaluates the cosine function. Hence, in contrast to finite-difference approximations, the derivatives obtained are accurate to working precision. 

Applying AD to the generating functions allows us to readily calculate all the quantities we have derived. \Crefrange{eq:multi_photon_detection_probability}{eq:multi_photon_rising_factorial_moments} as well as \cref{eq:photon_subtracted_generating_function,eq:photon_added_generating_function} only involve derivatives and basic functions as well as linear algebra that are easily implemented. With modern software tools, multivariate higher-order derivatives can be conveniently obtained with a few lines of code. From a practical point of view, this is convenient because the user is not confronted with the complexity of the calculation, which is hidden in the repeated derivatives. AD thereby greatly improves the practical usability of the generating functions we derived.
However, as mentioned in the introduction, the computational complexity to calculate the PND of GBS problems with state-of-the-art algorithms scales exponentially with the number of photons, so that the numerical evaluation of the derivatives of the generating function via AD will become infeasible for large photon numbers.

A variety of AD software exists and an overview of available tools is provided in \refcite{autodiff_org}. AD is widely used in machine learning, for example for training artificial neural networks~\cite{Baydin_2018}. Therefore, AD functionalities are part of popular machine learning libraries such as \href{https://pytorch.org}{\texttt{PyTorch}}~\cite{Paszke_2017,Paszke_2019} and \href{https://www.tensorflow.org}{\texttt{TensorFlow}}~\cite{tensorflow_2015}. Machine learning has gained considerable attention during the past years and the range of its applications is consistently growing. Consequently, software for machine learning will be further developed and it can therefore be expected that AD software will gain further capabilities during the next years, which of course is beneficial for our application. State-of-the-art machine learning libraries provide numerous possibilities to optimize their performance, for example, using parallel computing on graphic processing units~(GPUs).   

Throughout this paper, we use the \texttt{autograd}~function from the machine learning framework~\texttt{PyTorch 1.11.0} for AD. We use PyTorch in a very basic configuration, i.e.\ we do not use the option for acceleration by GPU computations and the only setting we change is the default precision which we increase from float32 to float64. All simulations are run on a regular desktop computer to show that for small photon numbers, the generating functions can be evaluated without much effort, demonstrating the practicability of our approach. In \refcite{Fitzke2022_Pytorch_Technical_Report}, we present a straightforward code example to demonstrate the use of PyTorch for computing the PND.

\section{Application to common Gaussian states}
\label{sec:Application_to_common_Gaussian_states}

In this section, we apply our simulation method to two common GSs with well-known photon statistics: the single-mode displaced squeezed thermal state and the two-mode squeezed vacuum with multiple squeezers. These practical examples demonstrate that our simulation method can easily take into account effects such as non-unity detection efficiencies, noise, and multiple modes entering the same detector.

\subsection{Photon number distribution of a single-mode displaced squeezed thermal state}

As a first example, we consider the single-mode displaced squeezed thermal state~\cite{Olivares_2012}
\begin{equation}\label{eq:displaced_squeezed_thermal_state}
    \hat \rho = D(\alpha)S(\chi)\hat \rho\ped{th}(\mu\ped{th})S^\dagger(\chi)D^\dagger(\alpha)
\end{equation}
with displacement operator~$D(\alpha) = \exp(\alpha \adag - \alpha^* \aope)$ and squeezing operator~$S(\chi) = \exp[(\chi\adag[2] - \chi^* \aope[2])/2]$. Here, $\chi= r\eh{\ii \theta}$ is the squeezing parameter and
\begin{equation}\label{eq:thermal_state}
    \hat \rho\ped{th}(\mu\ped{th}) = \sum_{m=0}^\infty \frac{{\mu\ped{th}}^m}{(1+\mu\ped{th})^{m+1}} \ket m \bra m
\end{equation} is a thermal state with mean photon number~$\mu\ped{th}$~\cite{Olivares_2012}.

By using \cref{eq:multi_photon_generating_function}, we  derive an analytic expression for the generating function\footnote{\label{fnt:DSTS_covariance}The covariance matrix~$\ms \Gamma$ for the state from \cref{eq:displaced_squeezed_thermal_state} is given by~\cite{Olivares_2012}
\begin{equation}\label{eq:Covariance_displaced_squeezed_thermal_state}
(1+2\mu\ped{th})\mleft[\mqty{1 & 0\\0 & 1}\cosh 2r +\amqty{r}{\cos \theta &\sin \theta \\\sin \theta  & -\cos \theta }\sinh 2r\mright]\nonumber
\end{equation}
and the displacement vector is $\vs d = \sqrt{2}\,[\Re(\alpha), \Im(\alpha)]\tran$.}:
\begin{equation}\label{eq:single_mode_analytic_G00z3}
G(w) = \frac{1}{\sqrt D}\exp\mleft(\frac{w E}{2D}\mright)\,.
\end{equation}
Here, we use the abbreviations
\begin{IEEEeqnarray}{rCl}
E &=& \avs{\alpha}[w-2-w(1+2\mu\ped{th})\cosh 2r]\nonumber\\
& & +\: w(1+2\mu\ped{th})\Re(\alpha^2\eh{-\ii \theta})\sinh 2r \qq{and}\\
D &=&  1 -w + w^2\mleft(\frac{1}{2}+\mu\ped{th}+\mu^2\ped{th}\mright)\nonumber\\
& &-\:(1+2\mu\ped{th})\mleft(\frac{w^2}{2}-w\mright)\cosh 2r\,.
\end{IEEEeqnarray}

While an explicit expression for the PND can be derived from this generating function, we do not present it here as it can be found in \refcite{Wang_2012_Photon_number} where it was derived by using a different approach. The photon statistics of the single-mode squeezed thermal state, displaced squeezed thermal state, and squeezed displaced thermal state have been studied in Refs.~\onlinecite{Marian_1992, Marian_1993}.

As a simplification, we now consider a single-mode displaced squeezed state~$\ket \psi  = D(\alpha)S(\chi)\vac$, i.e.\ with $\mu\ped{th}=0$, for which the PND is given by~\cite{Wang_2012_Photon_number}
\begin{IEEEeqnarray}{rCl}
p(n) &=& \exp\mleft(-\avs\alpha + \Re\bigl(\alpha^2\eh{-\ii\theta}\bigr)\tanh r\mright)\frac{\tanh^n r}{n!2^n \cosh r }\nonumber \\
&&\times\: \avs{H_n\mleft(\frac{\alpha\eh{-\ii \theta/2}\sinh r -\alpha^* \eh{\ii \theta /2}\cosh r}{\ii \sqrt{\sinh 2r }}\mright)}\,,\nonumber\\ \IEEElabel{eq:PND_Squeezed_coherent_state_analytic}
\end{IEEEeqnarray}
where~$H_n$ is the n-th~Hermite polynomial.

\begin{figure}[btp]
\includegraphics[width=\columnwidth]{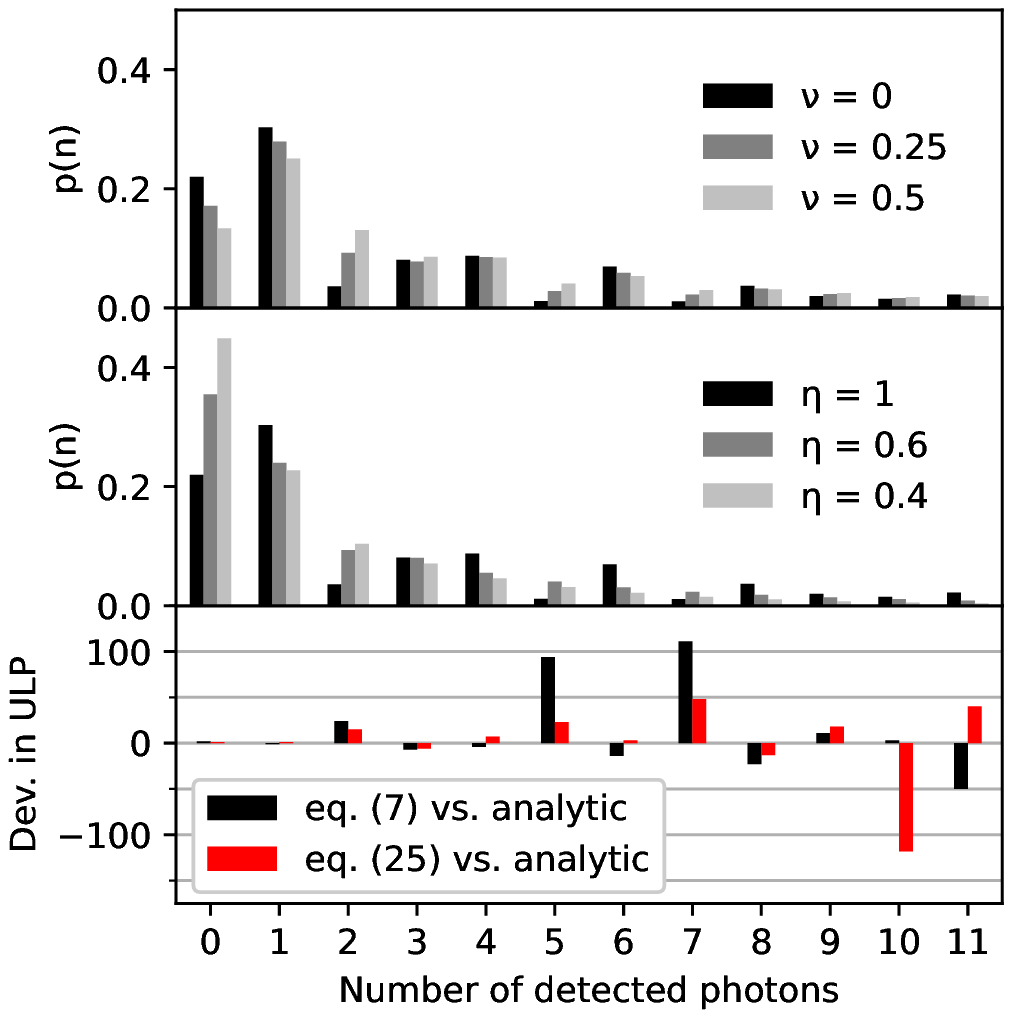}
\caption{Effect of Poissonian noise~$\nu$ and detection efficiency~$\eta$ on the detected photon number distribution of a displaced squeezed state~$D(\abs{\alpha}\eh{\ii \phi})S(r\eh{\ii\theta})\vac$ with $\avs{\alpha} = 1.2$, $\phi = \SI{50}{\degree}$, $\theta = \SI{30}{\degree}$, and $r\approx 1.287$ chosen such that~$\braket{\hat N} = 4$ by $\braket{\hat N} = \avs{\alpha} + \sinh^2(r)$~\cite{Gerry_2004}.
The upper diagram shows the results from \cref{eq:multi_photon_detection_probability} for different noise parameters~$\nu$ and the center diagram shows the influence of the efficiency~$\eta$. The bottom diagram shows the deviation in units of least precision (ULP) of float64 for the probabilities calculated from \cref{eq:multi_photon_detection_probability} and of the generating function \cref{eq:single_mode_analytic_G00z3} from the analytical value obtained via \cref{eq:PND_Squeezed_coherent_state_analytic}.
}
\label{fig:SMSS_statistics}
\end{figure}

In \cref{fig:SMSS_statistics}, we verify that deriving the PND of the state from the generating function \cref{eq:multi_photon_detection_probability,eq:single_mode_analytic_G00z3} for $\eta_{1/2} = 1$  and $\nu_{1/2} = 0$ by AD yields the same results as the analytic expression from \cref{eq:PND_Squeezed_coherent_state_analytic} within a tolerance of few units of least precision (ULP) of the analytical value. The maximum absolute value of the deviation from \cref{eq:multi_photon_detection_probability,eq:single_mode_analytic_G00z3} to the analytical result from \cref{eq:PND_Squeezed_coherent_state_analytic} is 118~ULP, i.e.\ the relative error of the numerical results is less than \num{2e-14} for both methods. This is remarkable because for $n=11$, computing the result via \cref{eq:multi_photon_detection_probability} already took a few minutes and required multiple~GB of~RAM, indicating a considerable number of numerical operations. The fact that the result is nevertheless numerically accurate underlines that the AD capabilities of \mbox{PyTorch} well suit our simulation method.

The analytic formula has the advantage that it can be quickly evaluated even for high photon numbers. But, an advantage of our formulation with a generating function in terms of $\ms \Lambda$~and~$\vs d$ is that effects such as noise and detection efficiencies can easily be taken into account (c.f.~\cref{fig:SMSS_statistics}).

\subsection{Multi-mode spontaneous parametric down-conversion}

A strength of our approach is that the detection statistics of states with multiple modes entering the same detector can be easily evaluated. For example, the spectral degree of freedom can be incorporated into a simulation by discretizing the frequency spectrum and treating each frequency as a separate mode. For the simulation in \cref{sec:Application_QKD_System} we want to take into account the photon statistics, but we do not resolve the frequency spectrum of the SPDC process. Because the spectrum and the photon statistics are related to each other, we recap some basic properties of the multi-mode SPDC photon statistics.

\begin{figure}[btp]
\includegraphics[width=\columnwidth]{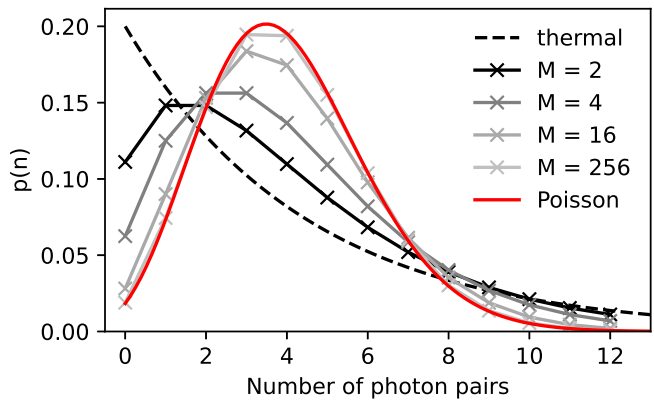}
\caption{Probability distribution of photon pairs produced by SPDC with mean photon number~$\mu = 4$ for different numbers of two-mode squeezers~$M$ with equal strengths. The distribution changes from thermal statistics for a single two-mode squeezer to Poissonian statistics for infinitely many two-mode squeezers. The distributions were obtained by AD of \cref{eq:Mauerer_analytic_Multimode_SPDC_genereating_function}.
\label{fig:Thermal_Poisson_Transition}}
\end{figure}

SPDC with one signal mode~$\mathrm s$ and one idler mode~$\mathrm i$ generates the two-mode squeezed vacuum~(TMSV) state $\ket{\psi}\ped{TMSV} = \exp(r\eh{\ii \theta} \adag\ped s\adag\ped i - r\eh{-\ii \theta} \aope\ped s \aope\ped i) \vac$ with mean photon number~$\mu = \sinh^2 r$ which can be written as~\cite{Wang_2008_Quantum_information}
\begin{equation}\label{eq:TMSV_state}
     \ket{\psi}\ped{TMSV} = \sech(r) \sum_{n=0}^\infty (\eh{\ii \theta}\tanh r )^n \ket{n\ped s} \ket{n\ped i}\,.
\end{equation} 
The photon statistics of either the signal or idler mode alone is therefore given by $p(n) = \sech^2(r) \tanh^{2n} r$, i.e.\ resembles that of a thermal state (cf.~\cref{eq:thermal_state}), with the probability generating function~$h(y) = \sech^2(r)/(1-y\tanh^2 r)$~\cite{Mauerer_2009}.
In practice, the bandwidths of the parametric process and the pump light result in two-mode squeezing between a number~$M$ of pairs of orthogonal Schmidt modes determined by the Schmidt decomposition of the joint spectral amplitude of signal and idler photons~\cite{Mauerer_2009,Thomas_2021}.
The resulting state is produced by applying $M$~independent two-mode squeezers to vacuum, each with thermal statistics of photon pairs. Hence, the generating function for the number of photon pairs is given by~\cite{Mauerer_2009}
\begin{equation}\label{eq:Mauerer_analytic_Multimode_SPDC_genereating_function}
    h(y) = \prod_{i=1}^M \frac{\sech^2 r_i }{1-y\tanh^2 r_i}
\end{equation} with $r_i \eh{\ii \theta_i} = C \sqrt{\lambda_i}/(\ii \hbar)$. The constant~$C$ is determined by the properties of the nonlinear medium and the pump light and~$\sqrt{\lambda_i}$ is the $i$-th Schmidt coeffcient.
For an increasing number of equally strong squeezers, the PND changes from thermal to Poissonian statistics~\cite{Mauerer_2009} as shown in \cref{fig:Thermal_Poisson_Transition}.

\begin{figure}[btp]
\includegraphics[width=\columnwidth]{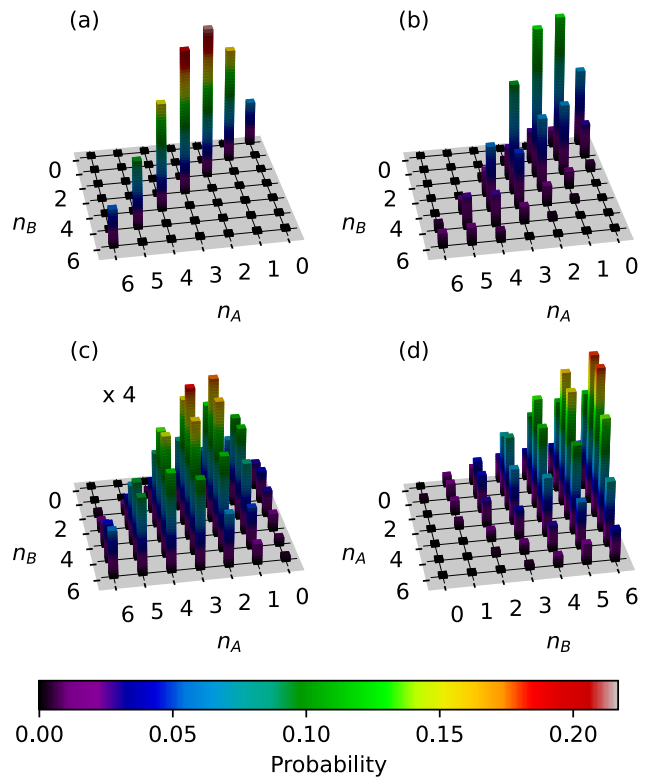}
\caption{Joint PND~$p(n_A, n_B)$ for a state with a mean photon pair number~$\mu = 3$ in~16 equally strong TMSV-modes with (a)~detection efficiencies $\eta_A= \eta_B = 1$, (b)~$\eta_A=\num{0.8}$ and $ \eta_B = \num{0.9}$ and (c)~efficiencies as in~(b)~and additional noise parameters $\nu_A = 1$~and~$\nu_B = 2$, scaled by a factor of~4 for better visibility. (d)~Cumulative PND $p(N_A = n_A, N_B \leq n_B)$ with noise and efficiencies as in~(b). \label{fig:joint_PND_TMSV}}
\end{figure}

As a further example for our simulation method, we consider the bivariate photon statistics of SPDC while taking into account noise and detection efficiencies. 
First, we calculate the two-dimensional joint distributions~$p(n_A, n_B)$ for different parameters of the detection efficiencies and noise as well as the cumulative distribution. The results obtained via our approach using the covariance matrix are shown in \cref{fig:joint_PND_TMSV}. It can be seen that as expected, decreasing the detection efficiency and adding noise blurs the line of delta functions that are observed for the ideal state along the diagonal~$n_A = n_B$.

\begin{figure}[btp]
\includegraphics[width=\columnwidth]{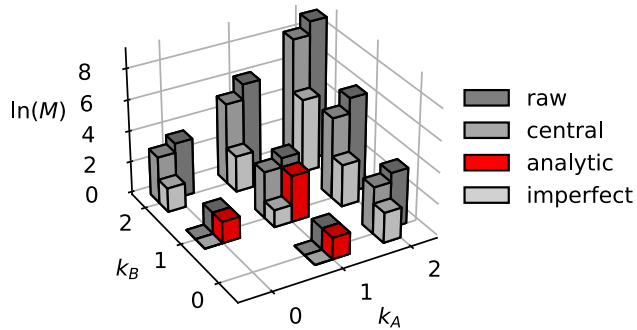}
\caption{Moments $M= \big\langle\big(\hat N_A -\braket{\hat N_A}\big)^{k_A}\big(\hat N_B-\braket{\hat N_B}\big)^{k_B}\big\rangle$ of the TMSV photon statistics with mean photon pair number~$\mu = 4$ calculated via \cref{eq:multi_photon_moments} in logarithmic scale. raw: raw moments; central: central moments; analytic: analytic expressions for the mean values $\braket{N_A} = \braket{N_B} = \sinh^2 r$ and for the first mixed moment $ \big\langle\big(\hat N_A -\braket{\hat N_A}\big)\big(\hat N_B-\braket{\hat N_B}\big)\big\rangle = \sinh^2(2r)/4$~\cite{Gerry_2004}; imperfect: central moments for an imperfect detector with $\eta_A = \num{0.5}$,  $\eta_B = \num{0.3}$,  $\nu_A = 1$~and~$\nu_B = 2$.
\label{fig:TMSV_mixed_moments}}
\end{figure}

Second, we use \cref{eq:multi_photon_moments} to analyze mixed moments of the PND of a TMSV state in \cref{fig:TMSV_mixed_moments}. For comparison, analytic values for the mean values and the first mixed moment are shown. However, if the photon statistics are more complicated e.g.\ due to loss, simultaneous detection of multiple modes, or noise, analytic expressions for the moments may not be easy to find, but can be easily calculated by applying AD to the moment-generating function \cref{eq:multi_photon_moments}. In \cref{fig:TMSV_mixed_moments}, we show the influence of loss and noise on the moments. 

To investigate the scaling of our method, we have measured the run time for the PND computation from \cref{fig:joint_PND_TMSV}(c) as a function of~$n_2$ with $n_1=0$ and for different numbers of modes. As expected for a GBS computation method, we observed an approximately exponential increase in the run time with the photon number. On average, each additional photon prolonged the run time by a factor of~\num{2.8} for 256~two-mode squeezers and~\num{3.5} for four two-mode squeezers. In comparison, Hafnian-based algorithms achieve a scaling of~$\mathcal{O}(N^3 2^{N/2})$ for $N$~photons~\cite{Bjoerklund_2019} and even faster methods have recently been demonstrated~\cite{Bulmer_2022}. It is important to note that the run time of our simulation method also grows with the number of modes. However, computing $p(n_1=0, n_2=6)$ for a $1024\times 1024$ covariance matrix representing 256 two-mode squeezers still took less than \SI{13}{\second}.

\section{Detection probabilities for entanglement-based Quantum Key Distribution}
\label{sec:Application_QKD_System}

For states and setups more complex than the examples considered above, analytical solutions can become infeasible and numerical simulations are required.
In this section, we consider an application demonstrating that by employing our method the photon statistics can be simulated in a relatively simple way, even for complex optical setups. We model a multi-user QKD system recently published based on the BBM92 protocol~\cite{Fitzke_2022_Scalable}. Although the system uses non-PNR detectors, the photon statistics are relevant as multi-photon pair emission is an important effect limiting the achievable quantum key rates. Detection efficiencies and noise are taken into account. Imperfect interference is modeled by using the mode mismatch model from \refcite{Takeoka_2015}. 
All these effects are readily included by using our simulation method.

\subsection{Simulated setup for quantum key distribution}

\begin{figure*}[btp]
\includegraphics[width=\textwidth]{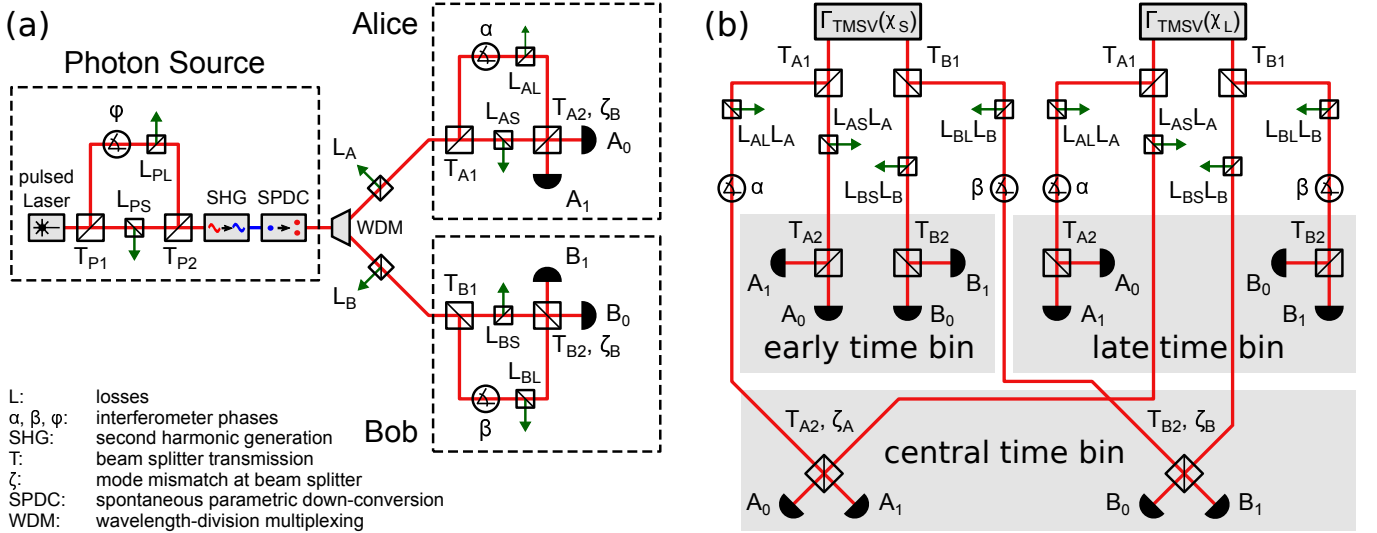}
\caption{(a)~Schematic setup of our BBM92 quantum key distribution system from \refcite{Fitzke_2022_Scalable} consisting of a central entangled photon pair source and receiver modules for two users Alice and Bob each comprising an imbalanced interferometer. Note that here  Mach-Zehnder interferometers are shown for clarity, but in practice, Michelson interferometers with Faraday mirrors are used instead~\cite{Fitzke_2022_Scalable}. 
Losses for all connections are modeled by introducing beam splitters coupling a fraction of the signal to an auxiliary vacuum mode and tracing this mode out. (b)~Unfolded setup for the simulation. The SPDC pumped with double pulses from the pump interferometer is modeled by setting up a pair of two-mode squeezed states with different squeezing parameters $\chi_S$~and~$\chi_L$. The different time bins are modeled as separate modes and for the early, central, and late times dedicated detectors are introduced. Imperfect interference at the beam splitters is taken into account by using the mode mismatch model from \refcite{Takeoka_2015}.
\label{fig:Setup}}
\end{figure*}

QKD enables the distribution of secure keys between users based on information-theoretic principles from quantum physics~\cite{Gisin_2002, Scarani_2009, Xu_2020}. The entanglement-based QKD system for four users named Alice~(A), Bob~(B), Charlie~(C), and Diana~(D)~\cite{Fitzke_2022_Scalable} implements the BBM92 time-bin protocol~\cite{BBM92, Brendel_1999, Tittel_2000, Marcikic_2004, Honjo_2008, Takesue_10}. \Cref{fig:Setup}(a) shows the setup for the users Alice and Bob. The setup for Charlie and Diana is identical, but with different values for parameters such as insertion losses and coupling ratios as well as different detector properties. 

The central component of the QKD system is a source of time-bin entangled photon pairs. Laser pulses are sent through an imbalanced interferometer, resulting in double pulses with a time and phase difference given by the interferometer's path length imbalance. The double pulses produce photon pairs by type-0 SPDC in a nonlinear crystal. The mean photon pair number per double pulse is $\mu \approx \num{0.034}$. The spectral width of the pump is much narrower than the photon frequency spectrum so that the photons are frequency-correlated: for a pump frequency~$2\omega_0$, energy conservation requires that the frequencies of the signal and idler photon sum up to $\omega_s+ \omega_i = 2\omega_0$. The photons are separated by wavelength
demultiplexing~(WDM). The phase matching function is almost constant over several dozen nanometers so that photons are produced in multiple pairs of \SI{100}{\giga\hertz}-wide WDM channels.
A pair of users obtain entangled photons only if their wavelength channels are symmetrically arranged around the center frequency~$\omega_0$.  Multiple user pairs can thereby exchange quantum keys simultaneously and independently and the user combinations can be reconfigured by changing the WDM channel assignments.

In the receiver station of each user, the photons pass an interferometer with a path length difference matched to that of the source's interferometer. They are therefore detected in one of three different time bins, depending on the path combination of the laser pulse and the photons in the interferometers. The quantum bit values in the time basis are obtained from the arrival time of the photons. When both photons are detected in the central time bin, two-photon Franson interference~\cite{Franson_1989} leads to emission from correlated interferometer outputs and the bit values in the phase basis are obtained from the detector numbers 0~and~1~\cite{Brendel_1999,Tittel_2000,Fitzke_2022_Scalable}.

However, various imperfections of the real setup lead to bit errors. For example, the interferometer delays may not be perfectly matched, the detectors have a non-zero dark count probability and the beam splitters are not ideal 50/50~splitters. The ratio of the number of bit errors to the number of measured bits is called quantum bit error rate~(QBER). The higher the QBER, the lower the secure key rate that can be extracted. A simulation of the system can help to quantify the contribution of the different effects to the QBER and to optimize the setup. A higher value of~$\mu$, for example, yields a higher emission probability for photon pairs per laser pulse, resulting in a higher bit rate. However, due to the statistical nature of SPDC, increasing~$\mu$ increases the probability of multi-pair emission and thereby leads to a higher QBER. Therefore, a simulation inevitably needs to take into account the effect of multi-pair emission.

For the simulation, we treat the three time bins as separate modes. The setup from \cref{fig:Setup}(a) can thus be unfolded as depicted in \cref{fig:Setup}(b). It is divided into four parts: state generation, propagation in fibers and interferometers, imperfect interference with mode-mismatch at the beam splitters $A_2$~and~$B_2$, and detection in three time bins. 
The state after the source interferometer is modeled by initializing two TMSV states with covariance matrices $\ms \Gamma\ped{TMSV}(\chi_S)$~and~$\ms \Gamma\ped{TMSV}(\chi_L)$, where the beam splitter asymmetries and losses in the arms of the source interferometer are absorbed into the squeezing parameters
\begin{IEEEeqnarray}{rCl}
\chi_S &=& r_0[T_{\mathrm P 1} (1-L\ped{PS})T_{\mathrm P 2}]^{1/2} \qq{and}\\
\chi_L &=& r_0\eh{\ii \phi} [(1-T_{\mathrm P 1})(1-L\ped{PL})(1-T_{\mathrm P 2})]^{1/2}\,.
\end{IEEEeqnarray}
The coefficient~$r_0$ is chosen such that both states together have a certain mean photon pair number~$\mu$. The propagation in the fibers and interferometers is represented by the state transformations for beam splitters, phase shifts, and losses and the interference described by the mode-mismatch model from \cite{Takeoka_2015} is also modeled by beam splitters. Although the setup is complex, the construction of the final covariance matrix is straightforward: it is a combination of only four basic building blocks, namely the covariance matrix of a TMSV state and the Gaussian state transformations for beam splitters, phase shifts, and losses. Combining the building blocks is simply achieved by multiplying the matrices representing them.
The representations of the most important Gaussian states and their transformations can be found e.g.\ in Refs.~\cite{Wang_2008_Quantum_information,Weedbrook_2012,Olivares_2012, Adesso_2014, Takeoka_2015} so that constructing a setup's model becomes a relatively simple task.

In \cref{fig:Thermal_Poisson_Transition}, we have shown the effect of the number of two-mode squeezers on the photon statistics. To obtain results representing the correct photon statistics, we would have to replicate the setup from \cref{fig:Setup}(b) in the simulation for the correct number of two-mode squeezers, each with its individual mean photon pair number. The Schmidt number~$K$ can be used to estimate the number of two-mode squeezers based on the shape of the joint spectral amplitude of the SPDC process~\cite{Mauerer_2009,Mauerer_2009_PhDThesis,Christ_2011}. For the simulation, we roughly estimate that~$K$ is in the order of magnitude of 100 and we assume equally strong two-mode squeezers resulting in almost Poissonian photon pair statistics. As the beam splitters and losses are applied to all Schmidt modes, the covariance matrix is block-diagonal with~$K$ identical blocks. Instead of directly calculating the determinant of this large matrix, we apply the rule for block determinants and raise the determinant of one block to the power of~$K=100$. 

In the experiment, we use single-photon avalanche detectors thoroughly analyzed in \refcite{Fitzke_2022_Timedependent}. Relevant parameters are the dead time~$\tau\ped{dead} \approx \SI{10}{\micro\second}$, the efficiency~$\eta \approx \SI{20}{\percent}$, the after-pulse probability of a few percent and the dark count rates in the range from~\SI{300}{\cps} to more than~\SI{3000}{\cps} depending on the detector. The flexibility of our simulation method allows us to include all these effects in the model as described in \cref{sec:QKD_detection}.

\subsection{Comparison of simulated key rates and error rates to measurements}

For the simulation, we have measured the values of all relevant experimental parameters, i.e.\ for the beam splitter transmissions, propagation losses in the fiber links, insertion losses of the components, and interferometer mode mismatches. Then, we used these values to determine the parameters of the simulation model in \cref{fig:Setup}(b). For the detection efficiencies, dark count rates, and afterpulse probabilities of our detectors we have included values obtained from tomographic measurements~\cite{Fitzke_2022_Timedependent} in the detection operators as described in \cref{sec:QKD_detection}. 

We have simulated the sifted key rates and quantum bit error rates by multiplying the source repetition rate with the detection probabilities for key bits and quantum bit errors in the time basis and phase basis. 
We compare the results with the measurements we have reported in~\refcite{Fitzke_2022_Scalable} to check the validity of our simulation. For these experiments, the mean photon pair number had been set to~$\mu = \num{0.034}$. For comparison, we consider the time basis QBER instead of the overall QBER because in the experiment the interferometer phases fluctuate and they are readjusted by tuning the interferometer temperatures as to minimize the phase basis QBER. Therefore, the phase basis QBER is not necessarily always at the minimum. Although it is possible to adjust the phases in the simulation to mimic phase fluctuations, we have chosen the phases for optimal interference and compare only the time QBER, which is unaffected by the phase fluctuations. 

\begin{figure}[btp]
\includegraphics[width=\columnwidth]{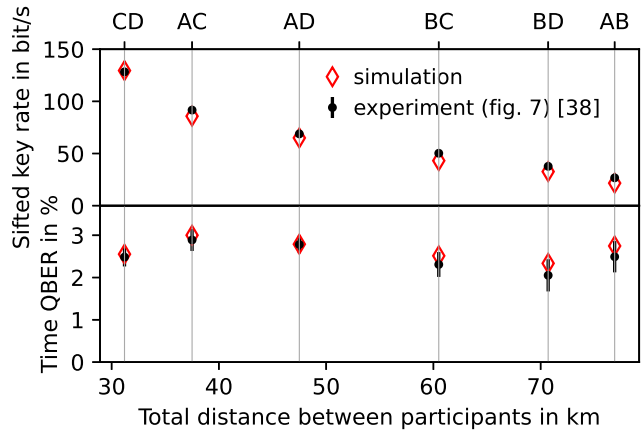}
\caption{Comparison of the experimental and simulated sifted key rates and quantum bit error rates~(QBERs) in the time basis for different transmission distances in the multi-user QKD network we have presented in Fig.~7 of \refcite{Fitzke_2022_Scalable}. The error bars show the standard deviation of the measured values over multiple runs. Different combinations of the four users Alice~(A), Bob~(B), Charlie~(C), and Diana~(D) exchange keys pairwise.
All parameters of the simulation such as the mean photon pair number per pulse, insertion losses, detection efficiencies, dark count rates, mode mismatches, and beam splitter coupling ratios were determined by measurements.
}
\label{fig:keyrate_QBER_comparision}
\end{figure}

A comparison between the measured and simulated sifted key rates and QBERs in the time basis is shown in \cref{fig:keyrate_QBER_comparision}. The simulation represents both the dependence of the key rates and of the time basis QBER on the user combination and distance very well. The key rate decreases with increasing distance between the participants due to the increasing propagation losses as expected. Since for the distances involved and the chosen pulse durations, dispersion is not significant, the time QBER is almost independent of the transmission distance and transmission losses. However, the QBER variations can be attributed to the individual dark count rates and afterpulse probabilities of the detectors. 
Multiple effects lead to quantum bit errors in the time basis: uncorrelated clicks from two noise counts can lead to coincidence. Or, one photon from a pair may be lost and the other photon can be detected in coincidence with a noise count or a photon from a different pair. Investigating the different contributions helps to determine the effects limiting the secure key rate and optimizing the setup. Thus, PNR simulations enable a detailed analysis of the different effects contributing to the QBER, as we demonstrate in \cref{sec:retrodictive_analysis}.

\subsection{Retrodictive analysis of the contribution of two-pair emission to coincidences}
\label{sec:retrodictive_analysis}

The process to derive the initial quantum state from the measurement outcome by applying Bayes' theorem is known as quantum retrodiction~\cite{Pegg_1999,Barnett_2000_Bayes,Barnett_2014,Barnett_2021} and has been applied for example to imperfect detectors with non\nobreakdash-unity quantum efficiency and dark counts~\cite{Barnett_1998} as well as to beam splitters and amplifiers~\cite{Jedrkiewicz_2004}.
Bayes' theorem states that the conditional probability~$p(E_1|E_2)$ for event~$E_1$, given that event~$E_2$ occurred, can be calculated from
\begin{equation}\label{eq:Bayes_Theorem}
    p(E_1|E_2) = \frac{p(E_2|E_1)p(E_1)}{p(E_2)}\,.
\end{equation}

The ability to simulate the photon number statistics enables such retrodictive analysis of the quantum state, i.e.\ it enables us to answer the question: Given that a certain measurement result was observed, what is the probability that it was produced by an initial quantum state with a specific photon number?

PNR simulations with GSs as presented above can be used to obtain~$p(E_1|E_2)$ when the probabilities $p(E_1)$~and~$p(E_2)$ can be obtained via the simulation and~$p(E_2|E_1)$ is also accessible. Such retrodictive analysis enables the investigation of effects that may not be easily accessible by measurements. For example, the influence of multi-photon pair emission on the QBER can be compared to the contribution from noise counts. Instead of simulating the probability for multi-photon pair production, it can be calculated for the photon statistics of the pair source or it can be estimated by measuring the heralded~$g^{(2)}$ auto-correlation function of the source~\cite{Montaut_2017}. Not all events in which multiple photon pairs are produced lead to quantum bit errors and conversely, not all quantum bit errors originate from multi-photon pair emission. Using PNR simulation, these effects can be included systematically. However, for example, the quantum bit errors originating from dark counts could also be estimated without setting up such a simulation.

When each of the two pump pulses produces a photon pair, Alice can detect a photon of the first pair in the early time bin and Bob can detect a photon of the second pair in the late time bin, causing a quantum bit error in the time basis. Here, we only want to briefly illustrate the method and for simplicity, we neglect the dependency of the time bins in the same repetition cycle due to the dead time among each other. Furthermore, we only consider the contribution from up to one photon pair produced by each pump pulse as well as raw coincidences between the detectors $A_0$~and~$B_0$ instead of all exclusive coincidences that would lead to error bits in the time basis. 

\begin{figure}[btp]
\includegraphics[width=\columnwidth]{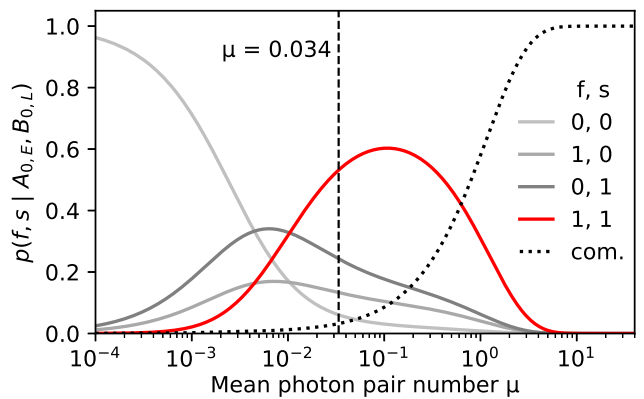}
\caption{Probability~$p(f,s |A_{0,\mathrm E}, B_{0,\mathrm L})$ that $f$~pairs were produced by the first pump pulse and $s$~pairs by the second pump pulse, given that a raw coincidence in $A_{0,\mathrm E}$~and~$B_{0,\mathrm L}$ was observed, for~$f, s \in \{0,1\}$. Here,~$\mu$ is the total produced mean number of photon pairs per repetition cycle over both pump pulses.  By 'com.' we denote the complementary probability, i.e.\ the difference of the sum of the other shown probabilities to one. It shows the contribution of pair numbers where  $f\ge2$~or~$s\geq2$ to the raw coincidence probability~$p(A_{0,\mathrm E}, B_{0,\mathrm L})$. The dashed line marks the working point $\mu = \num{0.034}$ in \refcite{Fitzke_2022_Scalable} (cf.~\cref{fig:keyrate_QBER_comparision}) with probabilities (from $f,s = 0,0$~to~$f,s=1,1$) of~\SIlist{6;13;24;53}{\percent}.}
\label{fig:retrodiction}
\end{figure}

Two photon pairs can only lead to an error in the time basis when one pair was produced by the first pump pulse happening with probability~$p(f)$, and the other one is produced by the second pulse occurring with probability~$p(s)$. Both values are obtained by simulating the probability to obtain one photon pair directly after the SPDC process.
From Bayes' theorem in~\cref{eq:Bayes_Theorem}, the probability can be calculated that when a coincidence between $A_{0,\mathrm E}$~and~$B_{0,\mathrm L}$ is measured, actually $f$ and $s$ pairs were produced by the first and second pulse, respectively:  
\begin{equation}\label{eq:f_s_given_two_clicks}
p(f,s |A_{0,\mathrm E}, B_{0,\mathrm L}) =  \frac{p(A_{0,\mathrm E}, B_{0,\mathrm L}|f,s )}{p(A_{0,\mathrm E}, B_{0,\mathrm L})}p(f,s)\,.
\end{equation}
The emission events are independent of each other, so that $p(f,s) = p(f)p(s)$. The detections are independent as well, so that in \cref{eq:f_s_given_two_clicks} we can factorize $p\big(A_{0,\mathrm E}, B_{0,\mathrm L}|f,s \big) = p(A_{0,\mathrm E}|f)p(B_{0,\mathrm L}|s)$. Here,~$p(A_{0,\mathrm E}|f)$ is the probability that detector~$A_0$ is triggered in the early time bin, given that $f$~pairs have been produced in the first pump pulse. For $f= 1$ it can be simply obtained by tracing the path of the photon through the setup to calculate the total photon transmission probability~$T_{A_{0,\mathrm E}}$ to the detector including the beam splitter transmissions, propagation losses, and the detector efficiency. The value for~$p(B_{0,\mathrm L}|s)$ is similarly obtained:
\begin{IEEEeqnarray}{lcrCllll}
p(A_{0,\mathrm E}&|&1) &=& p\ped{on,A_{0,\mathrm E}}&\big(1-\eh{-\nu_{A_{0,\mathrm E}}}&(1-T_{A_{0,\mathrm E}}&)\big)\,, \IEEElabel{eq:one_photon_transmission_probability_A}\\
p(B_{0,\mathrm L}&|&1)&=& p\ped{on,B_{0,\mathrm L}} &\big(1-\eh{-\nu_{B_{0,\mathrm L}}}&(1-T_{B_{0,\mathrm L}}&)\big)\,. \IEEElabel{eq:one_photon_transmission_probability_B}
\end{IEEEeqnarray}
For $f = 0$ and $s = 0$, the detection probability is given by the probability for a noise count, which is equivalent to setting $T=0$ in \cref{eq:one_photon_transmission_probability_A,eq:one_photon_transmission_probability_B}.
Finally, the raw time error coincidence probability~$p(A_{0,\mathrm E}, B_{0,\mathrm L})$ is obtained from the simulation as well.
Thereby, all probabilities can be calculated that are required to obtain~$p(f,s |A_{0,\mathrm E}, B_{0,\mathrm L})$ from \cref{eq:f_s_given_two_clicks}. In \cref{fig:retrodiction} we show~$p(f,s |A_{0,\mathrm E}, B_{0,\mathrm L})$ as a function of the produced total mean number of photon pairs per repetition cycle~$\mu$ and for different combinations of $f$~and~$s$. 

For small values of~$\mu$ below~\num{1e-3}, the largest contribution to the time basis error coincidence probability~$p(A_{0,\mathrm E}, B_{0,\mathrm L})$ comes from coincidences between dark counts. Coincidences involving one noise count and one photon pair are relevant for a wide range of $\mu$-values from $\mu<\num{1e-4}$~to~$\mu>\num{1}$. The difference between the cases $f, s = 1, 0$~and~$f, s = 0, 1$ is mainly due to different transmission losses from the fiber lengths of~\SI{26.9}{\kilo\meter} to Alice and~\SI{50.0}{\kilo\meter} to Bob and due to the much higher dark count probabilities for~$A_0$ of~\SI{3045}{\cps} compared to~$B_0$ with~\SI{606}{\cps}. The complementary probability in \cref{fig:retrodiction} shows that in the range around~$\mu \approx \num{0.1}$ and above, effects from more than one photon pair become relevant. In the range of~$\mu = \num{0.034}$ where the QKD setup is operated, a majority of \SI{53}{\percent} of the $A_{0,\mathrm E}, B_{0,\mathrm L}$-coincidences occur when each pump pulse produces one photon pair. This underpins the fact that a simulation of the QKD system should take into account effects from multiple photon pairs to produce accurate results for the QBERs.

\section{Discussion}
\label{sec:discussion}

In the previous sections, we have shown the versatile applications of our expressions for the calculation of the photon statistics of GSs. To the best of our knowledge, we are the first to derive expressions for cumulative probabilities, moments, and factorial moments for general multi-mode GSs.
For the PND~$p(\vs n)$, alternative calculation methods can be found in the literature. One approach requires $2n$~derivatives to evaluate the photon number~$n$. For that, $\ms \Gamma$~and~$\vs d$ are expressed in the complex $(\vaope, \vadag)$-basis instead of the real $(\vo x, \vo p)$-basis via $\ms \sigma _Q = \ms \Omega\dagg \ms \Gamma \ms \Omega/2+ \idM /2$ and $\vs \tau = \ms \Omega\dagg \vs d$. In our notation, the expression reads \cite{Hamilton_2017,Kruse_2019} 
\begin{IEEEeqnarray}{rCl}
p(\vs n) &=& \frac{1}{\vs n!\sqrt{\det \ms \sigma_Q }}\exp\mleft(-\frac{1}{2}\vs \tau \dagg \ms \sigma_Q^{-1}\vs \tau\mright)\nonumber\\
 & & \times\prod_i\mleft(\diffp{}{\alpha^{\vphantom *}_i, \alpha^*_i}\mright)^{n_i}\exp\mleft(\frac{1}{2}\vs \gamma \tran \ms A \vs \gamma + \vs \tau \dagg \ms \sigma_Q^{-1} \vs \gamma\mright)\IEEEeqnarraynumspace\IEEElabel{eq:PND_double_derivative}
\end{IEEEeqnarray}
with $\vs \gamma\tran = (\vs \alpha\tran, \vs \alpha\dagg)$ and
$\displaystyle \ms A = \mqty{\nullM & \idM_N \\ \idM_N & \nullM} (\idM_{2N}-\ms \sigma_Q^{-1})$.

The exponential to be differentiated has been recognized as the generating function for multivariate Hermite polynomials and therefore, the PND can be expressed in terms of these polynomials~\cite{Dodonov_1994, Kok_2001, Huh_2020}. However, these polynomials are given by complex recursion formulas and the expressions are therefore considered difficult to evaluate for higher numbers of photons~\cite{Berkowitz_1970,Cariolaro_2015,Phillips_2019}.

For GSs with $\vs d = \nullM$, \cref{eq:PND_double_derivative} can be rewritten in terms of the Hafnian function as
\begin{equation}\label{eq:hafnian_formula}
 p(\vs n) = \text{haf}(\ms A_S)/\big(n!\sqrt{\det \ms \sigma_Q}\big)\,,
\end{equation} 
where~$\ms A_S$ is derived from~$\ms A$ by repeating rows and columns, depending on the number of photons to be detected in a particular mode~\cite{Hamilton_2017,Quesada_2018,Kruse_2019,Bjoerklund_2019}. For states with $\vs d \neq \nullM$, a similar expression involves the loop Hafnian function~\cite{Quesada_2019_Simulating,Quesada_2019_FranckCondon,Bjoerklund_2019,Quesada_2022_Quadratic}. For non\nobreakdash-PNR detection, the probabilities involve the Torontonian and loop Torontonian function~\cite{Quesada_2018,Bulmer_2022_Threshold}. The software library \href{https://the-walrus.readthedocs.io/en/latest/index.html}{\texttt{The Walrus}}~\cite{Gupt_2019} provides software related to GBS and has implemented algorithms to evaluate these functions.
In contrast to such highly specialized algorithms, we use the general-purpose tool of AD for our computations.

Generating operators for the photon statistics are treated in textbooks such as Refs.~\onlinecite{Mandel_Wolf_1995,Barnett_2002_Methods}. In Refs.~\onlinecite{Mauerer_2009, Mauerer_2009_PhDThesis}, Mauerer~et~al.\ have applied AD to the generating function from \cref{eq:Mauerer_analytic_Multimode_SPDC_genereating_function} to obtain the PND resulting from multiple two-mode squeezers. In \refcite{Nauth_2022}, a probability-generating function for TMSV states in terms of the covariance matrix has been presented. However, generating functions in combination with AD, have rarely been applied for practical calculations in the context of photon statistics. A possible reason is that the evaluation of higher-order derivatives is in general a resource-intensive computational task so that analytical expressions are preferred. But, for the calculation of the photon statistics, this is not an a-priori disadvantage, as the required computational resources for GBS computations scale exponentially even with optimized state-of-the-art algorithms~\cite{Bjoerklund_2019,Bulmer_2022}. It can nevertheless be expected that specially tailored algorithms yield performance benefits becoming relevant for higher photon numbers.

Often only the lowest few moments of probability distributions, such as to order 3~or~4 are considered, which are easily evaluated by our method.
For the PND, we have been able to calculate photon numbers up to 10~to~12 without any optimizations. 
The simulations took a few minutes and required multiple~GB of~RAM. 
By tuning the parameters of PyTorch, by using different AD software or more powerful hardware, even higher numbers of photons can be simulated. In the field of GBS complexity research, PNDs for considerably higher numbers of photons have been computed on much more powerful computers. For example, GBS probabilities have been computed on a workstation with 96~CPUs for 50~photons in \refcite{Quesada_2022_Quadratic} or for up to 92~photons on a 100,000-core supercomputer in \refcite{Bulmer_2022}. 

However, quantum optics applications often consider states with low mean photon numbers, for example when coincidences are considered. An example is entanglement-based QKD where~$\mu\ll 1$ is required to avoid quantum bit errors from coincidences between photons from different pairs. The QKD system we have modeled in \cref{sec:Application_QKD_System} is operated with a mean photon pair number per pulse around~$\mu=\num{0.034}$  (cf.~\refcite{Fitzke_2022_Scalable}). For such low mean photon numbers, only the first few photon numbers are of practical relevance as the probabilities of finding higher photon numbers decrease rapidly. Thus, for many applications, a computational limitation to low photon numbers is not a significant restriction.

Another method to calculate the PND of GSs has recently been presented in~\refcite{Thomas_2021}, where one of the major results written in our notation is the expression
\begin{IEEEeqnarray}{rCl}
p(\vs n) &=& \eval{\frac{(\bnum{-1})^{\vs n}}{\vs n!}\vpdv[\vs n]{}{{\vs y}} \det\mleft(\idM+ \frac{\ms Y(\ms \Gamma -\idM)\ms Y}{2}\mright)^{- 1/2}}_{\abovebaseline[1pt]{ \scriptstyle \vs y = \bnum 1}} \IEEEeqnarraynumspace \IEEElabel{eq:Thomas_result}
\end{IEEEeqnarray}
with $\vs Y =\diag(\vs w) \oplus \diag(\vs w) $ and $w_s = \sqrt{y_j}$. The authors of \refcite{Thomas_2021} derive it by applying a fanning-out transformation to infinitely many virtual non\nobreakdash-PNR detectors, motivated by the intuition that in this limiting case, the probability that any of these virtual detectors receives more than one photon vanishes. Therefore, this approach formalizes the experimental approach to realize PNR detection by multiplexing multiple non\nobreakdash-PNR detectors mentioned in \cref{sec:introduction}.
The expresssion from \refcite{Thomas_2021}, given in \cref{eq:Thomas_result}, can be rewritten as a special case of our \cref{eq:multi_photon_detection_probability}, which generalizes the result from \refcite{Thomas_2021} in multiple ways: it already incorporates detection efficiencies and noise counts and it can be applied to states with non-zero displacement vector~$\vs d$.
Besides \cref{eq:multi_photon_detection_probability}, our derivation yields generating functions that enable the calculation of the moments, cumulative probabilities, density matrix elements, and factorial moments of the PND via \crefrange{eq:multi_photon_cumulative_probability}{eq:multi_photon_rising_factorial_moments} and \crefrange{eq:multi_coherent_matrix_element}{eq:multi_closeby_matrix_elements}. Furthermore, our expressions require, in contrast to \cref{eq:PND_double_derivative}, only one derivative per photon number facilitating the numerical evaluation.  While analytic expressions for the PND of single-mode photon-added and photon-subtracted GSs have been reported in~\refcite{Xu_2010,Wang_2012_Photon_subtracted,Wang_2012_Photon_number}, we have presented the more general \cref{eq:photon_subtracted_generating_function,eq:photon_added_generating_function} which cover the multi-mode case and which provide, besides the PND, also the cumulative probabilities, moments, and factorial moments.

Conceptually, our generating-function approach provides a new point of view on photon statistics that is different from the Hafnian approach closely related to graph theory, and also different from the approach using the fanning-out transformation. Therefore, it can be a starting point for further theoretical investigation. For example, the analytic generating functions could also be analyzed for the complexity of their computation. Another example is the analytical generating function we have presented for the single-mode displaced squeezed thermal state.

We have shown that the flexibility of our approach enables us to easily incorporate imperfections such as noise of arbitrary statistics, provided its generating functions are known.
In \refcite{Thomas_2021} it is noted that in comparison to other formulas for PNR detection, such as \cref{eq:PND_double_derivative} or \cref{eq:hafnian_formula}, an advantage of \cref{eq:Thomas_result} is that modes entering the same detector are treated by the same derivative, whereas the alternative formulas treat all modes separately so that terms for all the possible distribution patterns of the photons over the modes in a detector have to be calculated separately. In \refcite{Thomas_2021}, this fact has been used to incorporate the spectral modes into calculations of the Hong-Ou-Mandel visibility of SPDC sources. The same argument holds for our \crefrange{eq:multi_photon_detection_probability}{eq:multi_photon_rising_factorial_moments}, which involve one differentiation variable per detector, independent of the number of modes entering it. This is because our formulas obey the usual rules for generating functions for probabilities and moments and, therefore, the convolution of probability distributions is simply represented by the multiplication of the generating functions.

\section{Conclusions}

We have presented various generating functions for the photon statistics of multi-mode GSs in terms of the covariance matrix and displacement vector, from which the probabilities, moments, and factorial moments are retrieved by repeated AD. Common effects in experiments, such as noise, non\nobreakdash-unity detection efficiency, and the joint detection of multiple modes can be easily taken into account.
Furthermore, expressions for the generating functions of multi-mode photon-added and photon-subtracted GSs are derived, relating them to the generating functions of the underlying GSs, thus greatly extending the range of its possible applications.

We have implemented AD by using the software framework \mbox{PyTorch} to retrieve probabilities for photon numbers up to values of~12 on a desktop computer effortlessly, without any optimization for performance. The code for a simple example demonstrating the calculation of a PND using \mbox{PyTorch} can be found in our technical report \refcite{Fitzke2022_Pytorch_Technical_Report}.

As an application, we have modeled an entanglement-based system for QKD which we have recently developed and we have observed very good agreement between the simulated and measured sifted key rates and quantum bit error rates. 
Finally, we have used our PNR simulation of the setup 
to analyze the fraction of quantum bit errors due to the production of up to two photon pairs in the same repetition cycle. 

Our expressions for the generating functions in \crefrange{eq:multi_photon_detection_probability}{eq:multi_photon_rising_factorial_moments} are formulated in terms of simple operations on the covariance matrix and displacement vector of GSs. Together with AD for the differentiation of the generating functions, this enables practical, easy-to-implement, versatile simulations of quantum-optical setups. 

The data that support the findings of this study are available from the corresponding author upon request.

\section*{Acknowledgements}

This research has been funded by the Deutsche Forschungsgemeinschaft (DFG, German Research Foundation), under Grant No.\ SFB 1119--236615297. We thank Philipp Kleinpa\ss{} for fruitful discussions about the simulation of the QKD system. We also thank Julian K.\ Nauth, Vladimir M.\ Stojanović and Gernot Alber for valuable comments.

\appendix

\section{Basic properties of probability generating
functions and Gaussian states}

In this section, we briefly summarize basic properties of general multi-mode GSs and of generating functions for probability distributions.

\subsection{The covariance formalism for the simulation of Gaussian states}
\label{sec:Covariance_Formalism}

A photonic state with $S$~modes can be described by creation and annihilation operators~$\adag_s$ and~$\aope_s$ with~$s=1\dots S$. 
The quadrature operators~$\hat x_s = (\aope_s + \adag_s)/\sqrt 2$ and~$\hat p_s = (\aope_s - \adag_s)/\ii \sqrt 2$ can be expressed in matrix notation as~$\quadvo = \ms \Omega \ladvo$ with 
\begin{IEEEeqnarray}{rCl/c/r?l/c/lr}
\quadvo &=& \big(\hat x_1&\dots&\hat x_S&\hat p_1 &\dots& \hat p_S & \big) \tran \,, \IEEElabel{eq:def_quadrature_operators}\\
\ladvo &=& \big(\aope_1&\dots&\aope_S&\adag_1&\dots&\adag_S &\big)\tran\,,
\IEEElabel{eq:def_vector_creation_operators}
\end{IEEEeqnarray}
with the basis-changing matrix~$\displaystyle \ms \Omega = \frac{1}{\sqrt 2}  \amqty{r}{\idM & \idM  \\-\ii \idM & \ii\idM }$ and with $\idM$ denoting the $S\times S$~identity matrix.

The characteristic function of an operator~$\hat O$,
\begin{equation}\label{eq:characteristic_function_of_an_operator}
\chi_{\hat O} (\vs \xi ) =  \tr\bigl(\hat O \exp(\ii \vs \xi\tran \quadvo)\bigr)\,,
\end{equation} 
has $2 S$~arguments that can be collected into the vector $\vs \xi\tran = (\vs \xi_x\tran, \vs \xi_p\tran)= \mqty{\xi_{x_1}\dots\xi_{x_S}\, \xi_{p_1} \dots \xi_{p_S}}$. The expectation value $\braket{\hat O}$  w.r.t.\ a quantum state~$\hat \rho$ can be written as~\cite{Takeoka_2015, Wang_2008_Quantum_information}
\begin{equation}\label{eq:Expecation-value_from_characteristic_function_product}
\tr\bigl(\hat \rho \hat O\bigr) = \frac{1}{(2\pii)^S}\int_{\reals^{2S}}  \chi_{\hat \rho} (\vs \xi) \chi_{\hat O} (-\vs \xi) \dd{\vs \xi}.
\end{equation}

GSs are the states described by a Gaussian characteristic function~\cite{Wang_2008_Quantum_information,Weedbrook_2012,Olivares_2012,Takeoka_2015}
\begin{equation}\label{eq:characteristic_function_of_gaussian_state}
\chi_{\hat \rho} (\vs\xi) = \exp\biggl(-\frac{1}{4}\vs \xi\trans \ms \Gamma \vs \xi + \ii \vs d\trans \vs\xi\biggr)\, ,
\end{equation}
with a real, symmetric~$2 S \times 2S$ covariance matrix~$\ms \Gamma$ and a displacement vector~$\vs d$ defined by\footnote{We use the convention from  Refs.~\cite{Wang_2008_Quantum_information, Adesso_2014,Takeoka_2015}, as our notation benefits from grouping $x$~and~$p$ components together and the covariance of the vacuum states simply becomes~$\ms \Gamma = \idM$. The expressions we derive will depend on~$\ms \Gamma - \idM$. Other common conventions are to arrange $x$~and~$p$ in alternating order~\cite{Weedbrook_2012,Olivares_2012}, to scale~$\ms \Gamma$ by a factor of~1/2 \cite{Olivares_2012,Hamilton_2017,Kruse_2019}, or to define~$\ms \Gamma$ with complex elements with respect to~$\ladvo$ instead of~$\quadvo$~\cite{Adesso_2014,Kruse_2019,Hamilton_2017}.}
\begin{equation}
\Gamma_{ij} = \braket{\hat{\mathfrak q}_i\hat{\mathfrak q}_j + \hat{\mathfrak q}_j\hat{\mathfrak q}_i} - 2\braket{\hat{\mathfrak q}_i}\braket{\hat{\mathfrak q}_j}\qq{and} d_i =  \braket{\hat{\mathfrak q}_i}\,.
\end{equation}

Hamiltonians linear or quadratic in $\aope$~and~$\adag$ map GSs to GSs~\cite{Olivares_2012} and can be expressed in matrix notation as 
\begin{IEEEeqnarray}{rCllCl}
\hat H_1 &=& \ii \vs h \tran  \ladvo  &\qq{with} \vs h &=& \amqty{c}{-\vs \alpha^*\\ \vs \alpha} \qq{and} \\
\hat H_2 &=& \frac{1}{2}\ladvo \dagg \ms H \ladvo &\qq{with} \ms H  &=& \amqty{l}{\ms X & \ms Y \\ \ms Y^* & \ms X^*}\,.
\end{IEEEeqnarray}
Here,~$\ms Y= \ms Y \tran$ as well as~$\ms X= \ms X \dagg$ ensure that~$\hat H$ is Hermitian~\cite{Adesso_2014}. 
A unitary operation~$\hat U_{1,2} = \eh{-\ii \hat H_{1,2}}$ transforms~$\quadvo$ according to 
\begin{IEEEeqnarray}{rCl}
\hat U_1\dagg \quadvo \hat U^{\vphantom \dagg}_1  &=&  \quadvo +\ms \Omega \ms J \vs h =  \quadvo + \tilde{\vs d}\IEEElabel{eq:symplectic_quadrature_transformation_1}\,,\\
\hat U_2\dagg \quadvo \hat U^{\vphantom \dagg}_2 &=& \ms S \quadvo \qq{where} \ms S = \ms \Omega\eh{-\ii  \ms K \ms H}\ms \Omega\dagg\, \IEEElabel{eq:symplectic_quadrature_transformation_2}
\end{IEEEeqnarray}
as well as $\displaystyle \ms J = \amqty{r}{\nullM & \idM\\ -\idM & \nullM}$ and $\displaystyle \ms K = \amqty{r}{\idM & \nullM\\\nullM & -\idM}$.

The characteristic function becomes
\begin{equation}
\chi_{\hat \rho'} (\vs \xi ) =  \tr\bigl(\hat U \hat \rho \hat U\dagg \exp(\ii \vs \xi \tran \quadvo)\bigr) = \tr\bigl(\hat \rho \exp(\ii \vs \xi\tran  \hat U\dagg \quadvo \hat U )\bigr)\,,
\end{equation}
resulting in transformations of $\ms \Gamma$ and $\vs d$~\cite{Wang_2008_Quantum_information,Weedbrook_2012, Olivares_2012,Adesso_2014}:
\begin{IEEEeqnarray}{rCl"c"rCl?rCl}
\hat \rho' &=& \hat U^{\vphantom \dagg}_1 \hat \rho \hat U_1\dagg  & \Rightarrow & \ms \Gamma' &=& \ms \Gamma\,, & \vs d' &=& \vs d + \tilde{\vs d}\IEEEeqnarraynumspace\, \IEEElabel{eq:Gaussian_state_transformation_rule_1}\\
\hat \rho' &=& \hat U^{\vphantom \dagg}_2 \hat \rho \hat U_2\dagg & \Rightarrow & \ms \Gamma' &=& \ms S \ms \Gamma \ms S\tran\,, &\vs d' &=& \ms S \vs d \IEEElabel{eq:Gaussian_state_transformation_rule_2}
\end{IEEEeqnarray}

Examples of such transformations are the phase shift 
of a single mode and the coupling of two modes by a lossless beam splitter.

\subsection{Generating functions for probabilities and moments}
\label{ssec:Generating_functions}

The probability distribution of a non-negative integer-valued random variable~$N$ can be encoded by a probability generating function as~\cite{Chattamvelli_2019}
\begin{equation}\label{eq:PGF}
h(y) = \braket{y^N} =  \sum_{n=0}^\infty p(N = n)y^n
\end{equation}
so that the probabilities~$p(n)$ and the mean~$\braket{N}$ can be retrieved by
\begin{IEEEeqnarray}{rCl}
p(n) &=& \eval{\frac{1}{n!} \diff*[n]{h(y)}{y} }_{y=0} \qq{and} \IEEElabel{eq:PGF_mean_and_kth_raw_moment}\\
\braket{N} &=& \sum_{n=0}^\infty p(N = n) n = \eval{\diff*{h(y)}{y}}_{\abovebaseline[1pt]{\scriptstyle y=1}}\,.
\end{IEEEeqnarray} 
The series in \cref{eq:PGF} converges at least for~$y$ on the unit disk because all~$p(n)$ are non-negative and~$\sum_n p(n) = 1$. 

The cumulative probabilities~$p(N\le n)$ are encoded by a closely related generating function~\cite{Chattamvelli_2019}:
\begin{IEEEeqnarray}{rCl}
\frac{h(y)}{1-y} &=& \sum_{n,k=0}^\infty p(N = n)y^{n+k} \IEEElabel{eq:cumulative_PGF}\,,
 \\
p(N\le n) &=& \sum_{k=0}^n p(k)= \eval{ \frac{1}{n!}\diff*[n]{\frac{h(y)}{1-y}}{y}}_{\abovebaseline[1pt]{\scriptstyle y=0}}\,.
\end{IEEEeqnarray} 

The moments~${\cal M}(k,\mu) = \braket{(N-\mu)^k}$ of the probability distribution can be encoded in a moment generating function $M(\mu, y) = \braket{\exp[y(N-\mu)]}$~\cite{Chattamvelli_2019}:

\begin{IEEEeqnarray}{rCl}
M(\mu, y) &=& \sum_{n=0}^\infty p(n) \eh{y(n-\mu)} = \sum_{k=0}^\infty \frac{y^k}{k!} \braket{(N-\mu)^k}\,,\IEEEeqnarraynumspace\\
{\cal M}(k,\mu) &=& \eval{\diff*[k]{M_\mu(y)}{y}}_{\abovebaseline[1pt]{\scriptstyle y=0}}
\,.\end{IEEEeqnarray} 

For $\mu = 0$ or $\mu = \braket{N}$, $M(\mu, y)$ generates the raw moments~$\braket{N^k}$ or the central moments~$\braket{(N-\braket N)^k}$.

It is sometimes convenient to study factorial moments~\cite{Adam_1995, Mandel_Wolf_1995, Barnett_1998, Barnett_2018, Barnett_2002_Methods}. The falling factorial moments~$n_{(k)} = \braket{N_{(k)}}  = \braket{N(N-1)\cdots(N-k+1)}$ are generated
by $ \braket{(1+y)^N}$:
\begin{IEEEeqnarray}{rCl}
    h(1+y) &=&\sum_{n=0}^\infty p(n) \sum_{k=0}^n y^k\binom{n}{k} = \sum_{k=0}^\infty y^k\frac{\braket{N_{(k)}}}{k!} \IEEElabel{eq:falling_factorial_moment_gen_fun}\,, \IEEEeqnarraynumspace\\
    n_{(k)} & =& \eval{\diff*[k]{h(1+y)}{y}}_{\abovebaseline[1pt]{\scriptstyle y=0}}\IEEElabel{eq:falling_factorial_moment_retrieval}\,.
\end{IEEEeqnarray} 
Similarly, a generating function for the rising factorial moments~$n^{(k)} = \braket{N^{(k)}}  = \braket{(N+1)\cdots(N+k)}$ can be defined~\cite{Barnett_2018, Mandel_Wolf_1995}:
\begin{IEEEeqnarray}{rCl}
    R(y) &=&  \sum_{k=0}^\infty \frac{y^k}{k!} \sum_{n=0}^\infty p(n)\frac{(n+k)!}{n!} =\frac{1}{1-y} \Braket{(1-y)^{-N}}\,,\nonumber\\
    \IEEElabel{eq:rising_factorial_moment_gen_fun}\\
    n^{(k)} &= &\eval{\diff*[k]{R(y)}{y}}_{\abovebaseline[1pt]{\scriptstyle y=0}}\,. \IEEElabel{eq:rising_factorial_moment_retrieval} 
\end{IEEEeqnarray} 
In \cref{eq:rising_factorial_moment_gen_fun} the following relation is used~\cite{Wilf_2014}:
\begin{equation}
\frac{1}{(1-y)^{n+1}} = \sum_{k=0}^\infty \frac{(k+n)!}{k!n!} y^{k} \,. \label{eq:factorial_sum}
\end{equation}

In the derivation of the generating functions for multimode GSs, we use two important properties of such generating functions. First, multivariate probability distributions are encoded by multivariate generating functions. For example, the bivariate distribution~$p(n_1, n_2)$ is generated by~$h(y_1,y_2) = \sum_{n_1, n_2}p(n_1,n_2)y_1^{n_1}y_2^{n_2}$.

Second, the probability-generating or moment-generating function for the sum of random variables is the product of the probability-generating or moment-generating functions of the individual random variables~\cite{Chattamvelli_2019}. 
For example, the probability~$p(n)$ to detect $n$~photons in total, distributed over several modes can be calculated from the product of the generating functions~$h_m(y)$ of all individual modes, sharing the same parameter~$y$:
\begin{equation}\label{eq:Gen_fun_prod}
p(n) = \frac{1}{n!}\diff*[n]{\prod_{m} h_m(y)}{y} \bigg\rvert_{\abovebaseline[1pt]{\scriptstyle y=0}}\,.
\end{equation}

Note that when multiplying the generating functions for cumulative probabilities and falling factorial moments, the power series in~$N$ is multiplied by the prefactor $(1-y)^{-1}$ only once, not for each factor individually.

\section{Derivation of generating functions for the matrix elements and photon statistics}
\label{sec:Derivation_General_Generating_Function}

\subsection{Derivation of the relation between g(u,v,w) and matrix elements}
\label{app:Matrix_elemnts_singlemode}

In this section we derive expressions relating the matrix elements of~$\hat \rho$ to  $g(u, v, w) = \braket{\hat g(u, v, w)}$ from \cref{eq:general_generating_operator} for $\hat g(u, v, w) = \normOrd{\exp(u \aope +v \adag -w\adag \aope )}$.
We expand the quantum state in the overcomplete basis of coherent states~$\ket \alpha$ as ~$\hat \rho =  \int_\cplane P(\alpha)\ket \alpha \bra \alpha \dd[2]\alpha$ with the Glauber--Sudarshan P\nobreakdash-function~$P(\alpha)$ of the state. The expectation value of a normally ordered function of creation and annihilation operators~$ F(\adag, \aope)$ can be calculated by using the optical equivalence theorem~\cite{Mandel_Wolf_1995, Schleich_2001,Barnett_2002_Methods}
\begin{equation}\label{eq:optical_equivalence_theorem}
\tr(\hat \rho F(\adag, \aope)) = \int_\cplane P(\alpha)F(\alpha^*, \alpha)\dd[2]\alpha\,.
\end{equation}
Applying this result to the expectation value of~$\hat g$ yields
\begin{equation}\label{eq:phase_space_representation_of_g}
    \braket{\hat g(u, v, w)} = \int_\cplane P(\alpha) \eh{u \alpha + v\alpha^* - w\avs{\alpha}}\dd[2]{\alpha}\,.
\end{equation}
We obtain the matrix elements of~$\hat \rho$ in the basis of coherent states and number states:
\begin{IEEEeqnarray}{rCl}
    \braket{\alpha|\hat \rho|\beta} & = & \int_\cplane P(\gamma) \braket{\alpha|\gamma}\braket{\gamma|\beta} \dd[2]\gamma\nonumber \\
    &=&\eh{-(\avs{\alpha}+\avs{\beta})/2} \braket{\hat g(\alpha^*, \beta, 1)}, \IEEElabel{eq:coherent_state_matrix_element}\\
    \braket{n|\hat \rho|m} &=& \int_\cplane P(\alpha)\eh{-\avs{\alpha}} \frac{\alpha^n}{\sqrt{n!}}\frac{\alpha^{*m}}{\sqrt{m!}}\dd[2]{\alpha}\,.
\end{IEEEeqnarray}

Our goal is to evaluate the derivatives numerically. For $m = n+\Delta_m$ and $\Delta_m >0$, the number of derivatives can be reduced from $n+m$~to~$m$ by using
\begin{equation}
\braket{n|\hat \rho|n+\Delta_m} = \eval{\frac{(-1)^n}{\sqrt{n!m!}}
\diffp*[n,\Delta_m][m]{\braket{\hat g(0, v, w)}}{w,v}
}_{\abovebaseline[1pt]{\substack{v=0\\w=1}}}\,,
\label{eq:closeby_matrix_elements}
\end{equation} 
which is again directly evident from the phase space integral. 
Similar to \cref{eq:closeby_matrix_elements}, $\braket{m+\Delta_n|\hat \rho|m}$ can be calculated for $n>m$ by swapping the roles of $u$~and~$v$. \Cref{eq:closeby_matrix_elements} is especially useful for matrix elements close to the diagonal as it then considerably reduces the number of derivatives.

\subsection{Derivation of G(u,v,w) in terms of covariance matrix and displacement vector}
\label{app:Gen_fun_in_terms_of_covmat_and_disvec}

We now derive the generating function in terms of the covariance matrix and displacement vector. 
Inserting $\idO = \pii^{-1} \int_\cplane \ket \alpha \bra \alpha \dd[2]\alpha$ into  \cref{eq:optical_equivalence_theorem} allows us to calculate the trace of normally ordered operators via $\tr(F(\adag, \aope)) =\pii^{-1}\int_\cplane F(\alpha^*, \alpha)\dd[2]\alpha$.
By applying the Baker-Campbell-Haussdorf formula~\cite{Mandel_Wolf_1995} and cyclic permutation of the trace, we obtain the characteristic function from the trace of a normally-ordered operator by
\begin{equation}
\chi_{\hat g} (\vs\xi) = \tr\bigl(\hat g \eh{\ii (\xi_x \hat x +  \xi_p \hat p)}\bigr)= \eh{(\xi_x^2+\xi_p^2)/4}\tr\bigl(\eh{d\adag}\hat g\eh{c\aope}\bigr)
\end{equation} 
with~$c=(\ii \xi_x + \xi_p)/\sqrt{2}$ and~$d=(\ii \xi_x-\xi_p)/\sqrt{2}$. Separating the real and imaginary parts of~$\alpha = a + \ii b$ results in
\begin{IEEEeqnarray}{rCl}
\IEEEeqnarraymulticol{3}{l}{\tr\bigl(\eh{d\adag} \hat g \eh{c\aope}\bigr)\notag= \frac{1}{\pii}\int_\cplane \eh{d \alpha^*}\eh{u \alpha + v\alpha^* - w\avs{\alpha}} \eh{c \alpha}\dd[2]\alpha}\nonumber\\
\quad &=& \frac{1}{\pii}\int_\reals  \eh{(d+v+c + u)a-w a^2} \dd a \int_\reals  \eh{\ii (c+u-d-v)b-w b^2}\dd b\,.\nonumber\\
\end{IEEEeqnarray}
Evaluating the complex Gaussian integrals yields
\begin{IEEEeqnarray}{rCl}
	\chi_{\hat g} (\vs\xi) &=& \frac{1}{w}\exp\mleft(-\frac{1}{2}\frac{2-w}{2 w} \vs \xi\tran\vs \xi- \ii \vs \zeta\tran \vs \xi  +\frac{u v}{w}\mright)
	\:\:\text{with}  \IEEEeqnarraynumspace\IEEElabel{eq:SM_characteristic_function_h} \\
	\vs \zeta & =& \frac{1}{w\sqrt 2}\amqty{r}{-(u+v)\\ \ii (v-u)}\IEEElabel{eq:zeta_def}
\end{IEEEeqnarray} and  $\vs \xi \tran = (\xi_x, \xi_p)$.
By performing the above calculation for each mode separately, we generalize \cref{eq:SM_characteristic_function_h} to a multi-mode state with modes~$s = 1\dots S$ by defining~$Z = \sum_{s} u_s v_s w^{-1}_s$ and $\ms A = \diag\bigl(\frac{2- w_1}{2 w_1} ,\cdots,\frac{2-w_S}{2 w_S}\bigr)$ and extending $\vs \xi$~and~$\vs \zeta$ accordingly:
\begin{equation}
\chi_{\hat g} (\vs\xi) = \vs w^{\bnum{-1}}\exp\mleft(-\frac{1}{2}\vs \xi\trans (\ms A\oplus \ms A) \vs \xi - \ii \vs \zeta \tran \vs \xi + Z\mright)\,.
\end{equation}
Here, the direct sum $\ms A\oplus \ms A=\left(\begin{smallmatrix}\ms A & \nullM \\ \nullM & \ms A \end{smallmatrix} \right)$ is used to apply $\ms A$ to the $\vs  \xi_x$~and~$\vs \xi_p$ components in~$\vs \xi$.
By using \cref{eq:Expecation-value_from_characteristic_function_product,eq:characteristic_function_of_gaussian_state}, we calculate $G(\vs u, \vs v, \vs w) = \braket{\hat g(\vs u, \vs v, \vs w)}$ for the detection probabilities of a GS with characteristic function~$\chi_{\hat \rho} (\vs\xi) = \exp(-\vs \xi\trans \ms \Gamma \vs \xi/4 + \ii \vs d\trans \vs\xi)$
and obtain 
\begin{IEEEeqnarray}{rCl}
\IEEEeqnarraymulticol{3}{l}{G(\vs u, \vs v, \vs w) = \frac{1}{(2\pii)^{S}}\int_{\reals^{2S}}  \chi_{\hat \rho} (\vs \xi) \chi_{\hat g} (-\vs \xi) \dd{\vs \xi}} \nonumber\\
\quad &=& \frac{\vs w^{\bnum{-1}}}{(2\pii)^{S}} \int_{\reals^{2S}}   \exp\mleft(-\frac{1}{2}\vs \xi \tran \ms B \vs \xi+ \ii\vs z \trans \vs \xi+ Z\mright )\dd{ \vs \xi} \IEEElabel{eq:general_generating_function_W}\IEEEeqnarraynumspace 
\end{IEEEeqnarray} 
with $\ms B = \ms \Gamma / 2 + (\ms A\oplus \ms A) $ and $\vs z =\vs d + \vs \zeta$. We absorb the prefactor~$\vs w^{\bnum{-1}}$ into a diagonal matrix~$\ms W =  \diag(\vs w) \oplus \diag(\vs w)$, solve the integral 
\begin{equation}\label{eq:Gaussian_integral_determinant}
\int_{\reals^{2S}}\eh{-\vs \xi \trans \ms B \vs \xi/2 +\ii \vs z\tran \vs \xi} \dd{\vs \xi}
= \frac{(2\pii)^S}{\sqrt{\det \ms B}} \eh{- \vs z\trans \ms B^{-1} \vs z / 2}
\end{equation} and use $\vs w^{\bnum{-1}} = 1/ \sqrt{\det \ms W}$ and $\ms \Lambda  = \ms W \ms B$. Thereby  we obtain \cref{eq:multi_photon_generating_function}:
\begin{equation*}
G(\vs u, \vs v, \vs w) = \frac{\exp(-\vs z\trans \ms \Lambda^{-1} \ms W\vs z/2 + Z)}{\sqrt{\det \ms \Lambda}}\,.
\end{equation*}

\subsection{Derivation of the generating function for photon-added states}
\label{app:photon_added_derivation}

To derive expressions for the photon statistics of photon-added GSs, we consider $\tr(\hat \rho\aope[k] \hat g \adag[k]) $ and insert~$\idO = \sum_m \ket m \bra m$:
\begin{IEEEeqnarray}{rCl}
\tr\bigl(\hat \rho\aope[k] \hat g \adag[k]\bigr) 
&= &\sum_{n=0}^\infty\sum_{m =0}^\infty \braket{n|\hat \rho|m}\braket{m|\aope[k] \hat g \adag[k]|n}
\end{IEEEeqnarray}
Evaluating the expansion of $\hat g = \exp(-w \adag \aope)$ yields
\begin{equation}
\braket{m|\aope[k]\hat g\adag[k]|n}= 
(1-w)^{n+k}\frac{(n+k)!}{n!}  \delta_{mn}\,.
\end{equation} 
We insert the expression as a coefficient into a generating function,
apply the relation from \cref{eq:factorial_sum}
and insert $\braket{n|\hat \rho|n} = \braket{\normOrd{\exp(-\hat N)\hat N^n/n!}}$ from \cref{eq:quantum_mandel_formula}:
\begin{IEEEeqnarray}{rCl}
\IEEEeqnarraymulticol{3}{l}{\sum_{k=0}^\infty \frac{r^k}{k!}\tr\bigl(\hat \rho\aope[k] \hat g \adag[k]\bigr)
= \sum_{\mathclap{n,k=0}}^\infty\braket{n|\hat \rho|n} r^k(1-w)^{n+k}\frac{(n+k)!}{n!k!}}\nonumber\\
\quad &=& \sum_{n=0}^\infty \braket{n|\hat \rho|n}\frac{(1-w)^n}{(1-r(1-w))^{n+1}}\nonumber\\
&=& \frac{1}{1-r(1-w)} \,\bigg\langle{\normOrd{\exp\biggl(\frac{\hat N(1-w)}{1-r(1-w)}-\hat N\bigg)}\biggr\rangle}\,.
\end{IEEEeqnarray}

Finally, we obtain with $\tilde w = 1-[(1-w)^{-1}-r]^{-1}$
\begin{equation}
    \tr(\hat \rho\aope[k] \hat g \adag[k]) =\eval{\diff*[k]{\frac{G(0,0, \tilde w)}{1-r(1-w)}}{r}}_{\abovebaseline[1pt]{\scriptstyle r=0}}\,.
\end{equation}

\section{Model for photon detection in the QKD setup}
\label{sec:QKD_detection}

In general, a detector can be described by a positive operator-valued measure (POVM), i.e.\ a set of positive operators~$\{\hat \Pi_i\}$ with $\sum_i \hat \Pi_i = \idO$ so that the probability to obtain result~$i$ is $p(i) = \braket{\hat \Pi_i}$. Our detectors are non-PNR detectors, i.e.\ they can only yield two detection results: "click" and "no-click". 
In our QKD setup, we use the BBM92 protocol with time-bin entanglement. The detection events of the entangled photons are therefore cast in an early, central, and late time-bin (cf.~\cref{sec:Application_QKD_System}). Due to the dead time being much longer than the time bins and the time separating them, the detector cannot click in multiple time bins of the same repetition cycle. 
For our detectors, we have therefore four mutually exclusive results: Detection in the early~(E), central~(C), and late~(L) time bin as well as no~click~(no). The probability for a detector click in a particular interval is the probability that non-zero clicks are registered, multiplied by the probability~$p\ped{on}$ that the detector is not in the dead time. 
Hence, for the POVM elements we obtain $\idO = \hat \Pi \ped{E} + \hat \Pi \ped{C} + \hat \Pi \ped{L} + \hat \Pi \ped{no}$ with 
\begin{IEEEeqnarray}{rCl}
    \hat \Pi \ped{E} &=& p\ped{on}\idO\ped{C,L}\big(\idO\ped E-\hat \Pi_{N_E=0}\big)\,,\IEEElabel{eq:early_click_operator}\\
    \hat \Pi \ped{C} &=& p\ped{on}\idO\ped L\hat \Pi_{N\ped{E}=0}\big(\idO\ped C-\hat \Pi_{N\ped{C}=0}\big)\,,\IEEElabel{eq:central_click_operator}\\
    \hat \Pi \ped{L} &=& p\ped{on}\hat \Pi_{N\ped{E,C}=0}\big(\idO\ped L-\hat \Pi_{N\ped L=0}\big)\,,\IEEElabel{eq:late_click_operator}\\
     \hat \Pi \ped{no} &=& p\ped{off}\idO +p\ped{on}\hat \Pi_{N=0}\,.\IEEElabel{eq:no_click_operator}
\end{IEEEeqnarray} 
Here, we abbreviate the index~$N\ped{E,C,L} = N$, indicating all time bins together. The probability that the detector is live, i.e.\ not during its dead time~$\tau\ped{dead}$ at a given point in time, is given by $p\ped{on}= 1-p\ped{off}$.
The operators~$\hat \Pi_{N\ped{E}=0}$ in \cref{eq:central_click_operator} and~$\hat \Pi_{N\ped{E,C}=0}$ in \cref{eq:late_click_operator} take into account that due to the dead time, a click switches the detector off in the subsequent time bins of the same repetition cycle.

To include effects from afterpulses and dark counts occurring with the dark count rate~$r\ped{dark}$ and to calculate~$p\ped{on}$, we first consider the simple case of a detector showing some Poissonian noise~$\nu$ but exhibiting no dead time. We define the noise rate~$r\ped{noise} =  r\ped{dark} + p\ped{ap}r$, click rate~$r =  p_{\mathrm{click}}f\ped{rep}$ and click probability
\begin{IEEEeqnarray}{rCl}
p\ped{click}& = &\braket{\idO-\hat \Pi_{N=0}} = 1 - \frac{\exp(-\nu)}{\sqrt{\det \ms \Lambda}}\IEEElabel{eq:single_detector_click_probability}
\end{IEEEeqnarray}
with $\nu =r\ped{noise} / f\ped{rep}$. Here, $f\ped{rep}$ is the repetition rate of the photon pair source, i.e.\ $\nu$~and~$p\ped{click}$ are defined for a time interval of one repetition cycle.
The afterpulse probability~$p\ped{ap}$ is the probability that a click triggers a subsequent uncorrelated click.
Here, only modes of the covariance matrix are kept in~$\ms \Lambda$ which enter the detector, while the columns and rows corresponding to all other modes are removed. This means that on the one hand~$r\ped{noise}$ depends on the click rate and on the other hand the click rate also depends on the noise. A self-consistent value for~$r\ped{noise}$ can be obtained by solving
\begin{equation}\label{eq:darkcounts_including_afterpulses}
r\ped{noise}=  r\ped{dark} + p\ped{ap}f\ped{rep}\bigg(1 - \frac{\exp(-r\ped{noise}/f\ped{rep})}{\sqrt{\det \ms \Lambda}}\bigg)
\end{equation} for~$r\ped{noise}$. Linearizing the exponential for $r\ped{noise}/f\ped{rep} \ll 1$ yields
\begin{equation}\label{eq:nu_solved}
r\ped{noise} \approx\frac{r\ped{dark} + p\ped{ap}f\ped{rep}\mleft(1-1/\sqrt{\det \ms \Lambda}\mright)}{1-p\ped{ap}/\sqrt{\det \ms \Lambda}}\,.
\end{equation}
This approximation is justified because the values for~$r\ped{noise}/f\ped{rep}$ are in the order of~\num{1e-5} for our setup operated at values of~$\mu \approx \num{0.034}$. From \cref{eq:single_detector_click_probability}, we obtain the click rate $r$ including afterpulses. The noise parameter for detection in a time bin of width~$\Delta T= \SI{1}{\nano\second}$ is then given by~$\nu\ped{time\:bin} = r\ped{noise} \Delta T$.

Finally, we consider the effect of the dead time $\tau\ped{dead}$. Thus, we introduce the measured click rate~$r\ped m$ blocking a fraction~$p\ped{off} = r \ped m \tau\ped{dead}$ of the acquisition time. We can now calculate $p\ped{on} = 1- r\ped m \tau\ped{dead}$. Furthermore, the measured click rate is simply the product~$r\ped m = p\ped{on}r$ leading to 
$p\ped{on}  = 1/(1+r\tau\ped{dead})$, which can be directly calculated from the click rate.

For coincidences between detectors, we distinguish between raw and exclusive coincidences. By coincidences, we mean events where at least two detectors click in the same repetition cycle. While raw coincidences only consider that two detectors click, exclusive coincidences also require that the two other detectors do not click. For QKD, we use only those events for which an unambiguous bit value can be derived, i.e.\ only exclusive coincidences. Note that for the security of the key exchange, it is recommended that the participant observing clicks in both detectors randomly assigns one of the values~\cite{Luetkenhaus_2000,Xu_2020}, but for the sake of simplicity of the simulation, we work with exclusive coincidences which exclude such double-clicks. 
The exclusive coincidence probabilities, for example between detectors~$A_0$ in time bin $i$~and~$B_0$ in time bin~$j$ with $i, j \in \{E,C,L\}$, are given by 
\begin{IEEEeqnarray}{rCl}
p_{A_{0,i}, B_{0,j},\,\mathrm{excl.}} &=& \big\langle \hat \Pi_{A_{0,i}} \hat \Pi_{B_{0,j}}  \hat \Pi_{A_{1,\mathrm{no}}} 
\hat \Pi_{B_{1,\mathrm{no}}}
\big\rangle\,. \IEEElabel{eq:excl_coinc}
\end{IEEEeqnarray}

As each detection operator consists of two terms (cf.~\cref{eq:single_detector_click_probability}), expanding \cref{eq:excl_coinc} yields 16~terms.

To simplify the comparison of the simulation with the measured key rates, we approximate \cref{eq:no_click_operator}:
\begin{equation}\label{eq:no_click_approximation}
     \hat \Pi \ped{no} =\hat \Pi_{N=0}+ p\ped{off}(\idO - \hat \Pi_{N=0})
     \approx \hat \Pi_{N=0}\,.
\end{equation}
This approximation is valid when~$p\ped{off} \ll 1$ or when $\hat \Pi_{N=0}\approx \idO$, which is the case because the click probability is low due to the detection efficiency of $\eta=\SI{20}{\percent}$ and due to the transmission losses. 
Using this approximation reduces the number of terms from 16~to~4. As an example, consider exclusive coincidences between detector~$A_0$ in the early time bin and~$B_0$ in the late time bin. By using  \cref{eq:no_click_approximation}, the exclusive coincidence probability becomes
\begin{IEEEeqnarray}{rCl}
\IEEEeqnarraymulticol{3}{l}{p_{\mathrm{click}(A_{0,E}, B_{0,L}),\,\text{excl.}}} \nonumber\\
\quad &=& p_{A_{0,\mathrm{on}}}p_{B_{0,\mathrm{on}}}\Big\langle
\big(\idO_{A_{0,\mathrm E}}-\hat \Pi_{N(A_{0,\mathrm{E}})=0}\big)\nonumber\\
&&\times\: \hat \Pi_{N(B_{0,\mathrm{E,C}}A_1B_1)=0}\big(\idO_{B_{0,\mathrm{L}}}-\hat \Pi_{N(B_{0,\mathrm{L}})=0}\big) \Big\rangle\nonumber\\
&\approx& p_{A_{0,\mathrm{on}}}\,p_{B_{0,\mathrm{on}}}\exp(-\nu_{B_{0,\mathrm{E,C}}}-\nu_{A_1}-\nu_{B_1})\nonumber\\
&& \times\:\Biggl( 
\frac{1}{\sqrt{\det\ms \Lambda_{B_{0,\mathrm{E,C}}A_1B_1}}}
-\frac{\exp(-\nu_{A_{0,\mathrm E}})}{\sqrt{\det\ms \Lambda_{A_{0,\mathrm E}B_{0,\mathrm{E,C}}A_1B_1}}}\nonumber\\
&& -\:\frac{\exp(-\nu_{B_{0,\mathrm L}})}{\sqrt{\det\ms \Lambda_{B_0A_1B_1}}}+ \frac{\exp(-\nu_{A_{0,\mathrm E}}-\nu_{B_{0,\mathrm L} })}{\sqrt{\det\ms \Lambda_{A_{0,\mathrm E}B_0A_1B_1}}}\Biggr).\nonumber \\
\end{IEEEeqnarray}

In the experiment, the photons have been separated by WDM with an arrayed-waveguide grating, for which we have measured the wavelength-dependent transmission~$\tau(\omega)$ of Alice's and Bob's channels $C_A$~and~$C_B$ with channel widths~$\Delta$. In the simulation, we have included the averaged transmissions $\tau_{A/B} =\Delta^{-1} \int_{C_{A/B}}  \tau(\omega) \dd \omega$ into the losses $L_A$~and~$L_B$. Due to the photon frequency correlation, the average transmission probability for a photon pair is not $\tau_A \tau_B$ but  $\tau\ped{pair} =\Delta^{-1} \int_{C_A} \tau(\omega) \tau(2\omega_0 - \omega)\dd \omega$. We, therefore, multiply the key rate by the correction factor~$\tau\ped{pair}/(\tau_A \tau_B) \approx \num{1.5}$ to take the spectral dependence of the channel loss into account. 

\section*{References}

\providecommand{\noopsort}[1]{}\providecommand{\singleletter}[1]{#1}%

\end{document}